\documentclass{imsart}

\usepackage[OT1]{fontenc}
\usepackage{amsthm,amsmath,amssymb}
\usepackage{graphicx}
\usepackage{multirow,multicol}
\usepackage{array}
\usepackage{booktabs}
\usepackage{indentfirst}
\usepackage{chngcntr}
\usepackage[numbers]{natbib}
\usepackage[colorlinks,citecolor=blue,urlcolor=blue]{hyperref}
\usepackage{epsfig}
\usepackage{pdfpages}

\startlocaldefs
\counterwithin{figure}{section}
\counterwithin{table}{section}
\numberwithin{equation}{section}
\newtheorem{thm}{Theorem}[section]
\newtheorem{cor}[thm]{Corollary}
\newtheorem{lem}[thm]{Lemma}
\newtheorem{prop}[thm]{Proposition}
\theoremstyle{definition}
\newtheorem{defn}[thm]{Definition}
\theoremstyle{remark}

\newtheorem*{acc*}{}
\theoremstyle{plain}
\endlocaldefs

\allowdisplaybreaks

\begin{document}

\begin{frontmatter}

\title{Local Neighborhood Fusion in Locally Constant Gaussian Graphical Models}
\runtitle{Local NFL in Locally Constant GGM}
\thankstext{T1}{Funding for this work was provided by NSF grant DMS-1107206.}

\begin{aug}
\author{\fnms{Apratim} \snm{Ganguly}
\ead[label=e1]{apratimganguly@gmail.com}}
\ead[label=u1,url]{http://math.bu.edu/people/apratim}
\and
\author{\fnms{Wolfgang} \snm{Polonik}
\ead[label=e2]{wpolonik@ucdavis.edu}}
\ead[label=u2,url]{http://anson.ucdavis.edu/~polonik}

\runauthor{Ganguly, A. \& Polonik, W.}

\affiliation{University of California, Davis}

\address{Apratim Ganguly\\
Postdoctoral Associate\\
Department of Mathematics and Statistics\\
Boston University\\
111 Cummington Mall\\
Boston, MA 02215\\
\printead{e1}\\
\printead{u1}}

\address{Wolfgang Polonik\\
Professor\\
Department of Statistics\\
University of California, Davis\\
One Shields Ave.\\
Davis, CA 95616\\
\printead{e2}\\
\printead{u2}}
\end{aug}

\begin{abstract}
\noindent In this paper we penetrate and extend the notion of local constancy in graphical models that has been introduced by Honorio et al. (2009). We propose \emph{Neighborhood-Fused Lasso}, a method for model selection in high-dimensional graphical models, leveraging locality information. Our approach is based on an extension of the idea of node-wise regression (Meinshausen-B\"{u}hlmann, 2006) by adding a fusion penalty. We propose a fast numerical algorithm for our approach, and provide theoretical and numerical evidence for the fact that our methodology outperforms related approaches that are ignoring the locality information. We further investigate the compatibility issues in our proposed methodology and derive bound for the quadratic prediction error and $l_1$-bounds on the estimated coefficients.
\end{abstract}



\end{frontmatter}

\section{Introduction}\label{intro}
In the context of a graphical model, the concept of local constancy has been coined recently in Honorio et al., 2009. The goal of the present paper is to penetrate this idea by providing a thorough analysis and an extension of this concept. Our approach is using the idea propounded by Meinshausen and B\"{u}hlmann (2006) for structure learning in Gaussian graphical models, and it incorporates an additional fusion penalty term that aims to enforce the structural constraint of {\em local constancy}. When the nodes of a graph have spatial information attached, then local constancy can be interpreted as a certain type of spatial regularity (see below). Conceptually, the entire neighborhood graph given by the graphical model is split into two subgraphs: (a) A \emph{local} graph consisting of nodes that are spatially close, and (b) a \emph{non-local} graph, where the edges connect nodes of two different spatial clusters of nodes. Our approach then assumes that we have prior structural knowledge about the local graph that we aim to incorporate into the estimation/model selection approach.

It is well known that a $p$-dimensional graphical model is given by a $p$-dimensional Gaussian distribution with non-singular covariance matrix $\Sigma$ of a random vextor $X$. The conditional independence of the components of $X$ can be represented by occurrence of zero entries of precision matrix $\Sigma^{-1} = \Omega$. The non-zero entries of the precision matrix define the edges of the graph  corresponding to this multivariate Gaussian distribution. 

Neighborhood selection algorithms aim to find all the neighbors of a node $X_{a}$ in a graphical model based on an i.i.d. sample. Meinshausen \& B\"{u}hlmann~\cite{Meinshausen2006} showed that this problem can be interpreted as an ensemble of $l_1$-penalized regressions, each of which can be solved using the lasso algorithm of Tibshirani (1997). Our proposal is an extension of this approach. In order to incorporate this additional structural constraint we assume that the prior structural knowledge can be translated into a \emph{local} neighborhood graph, which we think of being comprised of regular graph objects such as chains, cycles, lattices or cliques. The availability of additional locality information is critical when data is observed in a certain manifold with spatial geometry. The assumption of local constancy, in some sense, enforces spatial regularization on structure learning and thereby stimulates the search for probabilistic dependencies between local clusters of nodes. We would like to emphasize that the knowledge of this prior structural information is based on domain knowledge and hence known beforehand.

We propose to use the Meinshausen \& B\"{u}hlmann approach and to add a new penalty term which, in essence, generalizes the fused lasso penalty (see Tibshirani, 2005) by extending the prenalty term over differences of nodes in the local neighborhood graph, which in turn is given by the prior structural information. This leads to the \emph{neighborhood-fused lasso} procedure, a model selection method for locally constant Gaussian graphical models. We provide theoretical and numerical evidence that our approach outperforms competing model selection algorithms where locality information is ignored. 

The paper is organized as follows: In section~\ref{review} we first provide an overview of related work. Sections~\ref{NFL} and~\ref{LocConstDefn} then provide a precise definition of our approach descibed above, and a discussion of local constancy, respectively.  In section \ref{optimization} we propose our optimization algorithm, before we present both both finite sample and large sample properties of our appraoch in section~\ref{theory}. We prove theoretically that introducing a local penalty term reduces the finite sample type-I error probability in model selection and leads to equivalent accuracy with smaller sample size than competitors. We also discuss data dependent choice of the penalty parameters, which avoids the use of cross-validation based methods. We study the asymptotic $l_1$ properties of our estimator to find sufficient conditions on design matrix and regularization parameters to find nice asymptotic bounds in parameter estimation and prediction in terms of $l_1$ and $l_2$ metric, respectively.  In particular, our theoretical analysis reveals that with our assumptions our proposed method displays all the desired properties like sign-consistency and model selection consistency. Some numerical results can be found in section~\ref{simulations}. The proofs are delegated to the appendix.

\section{A Review of Related Work}\label{review}
Going back to Dempster~\cite{Dempster1972} who introduced \emph{Covariance Selection} to discover the conditional independence restrictions (the graph) from a set of i.i.d. observations, many methods have been proposed for sparse estimation of the precision matrix in a Gaussian graphical model. The proposed procedures usually rely on optimization of an objective function~\cite{Edwards2000,Lauritzen1996}. While modern technological developments and high computing power enable us to deal with high dimensional models, there still are computational challenges. Usually, greedy forward-selection or backward-deletion search is used. In forward (backward) search, one starts with the empty (full) set and adds (deletes) edges iteratively until a suitable stopping criterion is fulfilled. The selection (deletion) of an edge requires an MLE fit~\cite{Speed1986} for $O(p^2)$ many models, making it a suboptimal choice for high-dimensional models where $p$ is large. Also, the MLE might not exist in general for $p>n$ (see \cite{Buhl1993}). In contrast, neighborhood selection using lasso, as proposed by Meinshausen and B\"{u}hlmann~\cite{Meinshausen2006}, relies on optimizing a convex function applied consecutively to each node in the graph, thus fitting $O(p)$ many models. Fast lasso-type algorithms and data dependent choices for regularization parameter reduce the computational cost. Unlike covariance selection, this algorithm estimates the dependency graph by sequential estimation of individual neighbors and subsequent combination by taking unions or intersections. Other authors have proposed algorithms for the exact maximization of the $l_1$-penalized log-likelihood. Yuan \& Lin~\cite{Yuan2007} proposed an $l_1$-penalty on the off-diagonal elements of the concentration matrix for its sparse estimation with the positive definiteness constraint. They showed that this problem is similar to the \emph{maxdet} problem (see Vandenberghe et al.~\cite{Vandenberghe1998}), and thus solvable by the interior point algorithm. A quadratic approximation to the objective function in their proposed method leads to a solution similar to Meinshausen \& B\"{u}hlmann. Banerjee et al.~\cite{Banerjee2008} viewed this as a penalized maximum likelihood estimation problem with the same $l_1$-penalty on the concentration matrix. Constructing the dual transforms the problem into sparse estimation of the covariance matrix instead of the concentration matrix. They proposed \emph{block coordinate descent algorithm} to solve this efficiently for large values of $p$. They also showed that the dual of the quadratic objective function in the block-coordinate step can be interpreted as a recursive $l_1$-penalized least square solution. Friedman, Hastie \& Tibshirani~\cite{Friedman2007} used this idea successfully to develop an algorithm known as \emph{graphical lasso}. Using a coordinate descent approach to solve the lasso problem speeds up the algorithm to a considerable extent, making it quite fast and effective for a large class of high-dimensional problems.

In all the aforementioned methods, information on the local geometry is not taken into consideration. Often one encounters data that are measured on a certain manifold. E.g., data describing some feature on the outline of a (moving) silhouette, or pixels or voxels in an 2-d or 3-d image, respectively. In most of these problems, spatially close variables have a structural resemblance in terms of probabilistic dependence. Exploiting this local behavior might lead to faster and/or more efficient estimation. Honorio, Ortiz, Samaras et al.~\cite{Honorio2009} introduced the notion of \emph{local constancy}. According to them, if one variable $X_a$ is conditionally (in)dependent of $X_b$, then a local neighbor of $X_a$ in that manifold is likely to be conditionally (in)dependent of $X_b$ as well. They developed a\emph{coordinate direction descent algorithm} to solve a penalized MLE problem that penalizes both the $l_1$-norm of the precision matrix and the $l_1$-norm of local differences of the precision matrix, expressed as its ``diagonal excluded product" with a local difference matrix. Local geometry has been addressed, although implicitly, by Tibshirani et al.~\cite{Tibshirani2005} in the context of fused lasso. Chen et al.~\cite{Chen2010} used this idea and proposed \emph{graph guided fused lasso} for structure learning in multi-task regression, where the output space is continuous and the outputs are related by a graph. The input space is high dimensional and outputs that are connected by an edge in the graph are believed to share a common set of inputs. The goal then is to learn the underlying functional map from the input space to the output space in a way that respects the similar sparsity pattern among the covariates that are believed to affect the output variables that are connected. Local smoothing by penalizing differences of neighboring nodes has been discussed in Kovac and Smith~\cite{Kovac2012} in the context of nonparametric regression based on observations on nodes of a graph. Their algorithm aims to split the image into active sets and subsequently merge, split or amalgamate them in order to minimize a penalized weighted distance.

\section{Neighborhood selection using Fused Lasso}\label{NFL}
Like Greenshtein \& Ritov~\cite{Greenshtein2004} and Meinshausen \& B\"{u}hlmann~\cite{Meinshausen2006} we work in a set-up where the number of nodes in the graph $p(n)$ and the covariance matrix $\Sigma (n)$ depend on the sample size. Consider the $p(n)$-dimensional multivariate random variable $X=(X_1,\cdots,X_p)\sim N(\mu,\Sigma)$. The conditional (in)dependence structure of this distribution can be represented by the graph $\mathcal{G}=(\Gamma(n),E(n))$, where $\Gamma(n)=\{1,\cdots,p(n)\}$ is the set of nodes corresponding to each coordinate variable and $E(n)$ the set of edges in $\Gamma(n) \times \Gamma(n)$. A pair of nodes $(a, b)$ lies in $E(n)$ if and only if $X_a$ is conditionally dependent of $X_b$, given all other remaining variables $X_{\Gamma(n)\backslash\{a,b\}} = \{X_k : k \in \Gamma(n)\backslash\{a,b\}\}$. 

The neighborhood $\mathrm{ne}_{a}$ of a node $a$ is defined as the smallest subset of $\Gamma(n)\setminus\{a\}$ such that given $\mathrm{ne}_{a}$, $X_{a}$ is conditionally independent of all the remaining nodes. In other words, neighbors of a certain node consists of the coordinates that are conditionally dependent on that particular node. As already mentioned,  the conditional independence can be represented by occurrences of zero entries at respective cells of precision matrix $\Omega$, i.e., $X_a \perp X_b | X_{\Gamma(n)\setminus\{a,b\}}$ iff $\Omega_{ab} = \Sigma^{-1}_{ab}=0$. The neighborhood selection / model selection algorithms aim to find all the neighbors of a node $X_{a}$, given $n$ i.i.d. observations of $X$. Meinshausen \& B\"{u}hlmann considered this as a penalized regression problem, where each variable is regressed on the remaining variables with an $l_1$ penalty on the estimated coefficients. 

By the above definition of neighborhood, we have, for all $a\in \Gamma(n)$ that
\begin{align*}
X_{a}\quad &\bot \quad \big\{X_{k}:k\in \Gamma(n)\setminus \{a\cup\mathrm{ne}(a)\}\big\} \;\; | \;\; X_{\mathrm{ne}(a)}.
\end{align*}
An alternative definition of neighborhood of $X_a$ is given as the non-zero components of $\theta^{a}$ where $\theta^{a}$ is given by
\begin{align}\label{thetaregdefn}
\theta^{a} \; &=\;\mathrm{argmin}_{\theta: \theta_{a}=0}\mathbb{E}\left(X_{a}-\sum_{k\in \Gamma}\theta_{k}X_{k}\right)^2.
\end{align}
In light of the above definition, the set of neighbors of a node $a\in \Gamma(n)$ is
\begin{align}\label{nbddefn}
\mathrm{ne}(a)=\{b\in \Gamma(n): \theta^{a}_{b}\ne 0\}.
\end{align}
Given the domain knowledge we construct a regular graph ${\cal G}_{\rm local}$ that is representative of the underlying spatial geometry. We call it a \emph{local neighborhood graph}. For example, ${\cal G}_{\rm local}$ could be a chain graph, a two or three dimensional lattice or more generally a collection of cliques or wheels. Let $E_{\rm local}$ denote the edge set corresponding to ${\cal G}_{\rm local}$, then define:
\begin{defn}\label{localgraphdefn}
$X_a$ and $X_b$ are \emph{local neighbors} with respect to ${\cal G}_{\rm local}$ if the edge connecting them belongs to $E_{\rm local}.$
\end{defn}
Note that the edge between $X_a$ and $X_b$ need not belong to $E(n)$. Now if one incorporates the \emph{local constancy} property to this graph, the neighborhood becomes more structured. By local constancy, if $X_{b}$ is conditionally independent of $X_{a}$ given the other nodes (and hence, there is no edge between $a$ and $b$), it is likely that a {\em local neighbor} $X_{b'}$ of $X_{b}$ is also conditionally independent of $X_a$, making both $\theta^{a}_{b}$ and $\theta^{a}_{b'}$ equal to zero. Thus, the zeroes of $\theta^{a}$ are expected to reflect the sparsity pattern.

We now generalize the traditional fused lasso algorithm~\cite{Tibshirani2005} to satisfy our purpose of estimating the zeros and non-zeros of the precision matrix by exploiting the locality information given by the local neighborhood graph ${\cal G}_{\rm local}$. Before doing that we carefully define the \emph{difference matrix}. Consider an $m\times p$ matrix $D$ where $m = |E_{\rm local}|$ is the total number of pairs of local neighbors. Assuming that we have a labelled sequence of nodes $\Gamma(n) = \{1,2,\cdots,p(n)\}$, arrange the pairs of local neighbors in a sequence
\begin{align*}
\mathcal{B} := \left\{\left\{(u,v_u): v_u\in \mathrm{ne}_{l}(X_u) > u\right\}: u\in \Gamma(n)\right\},
\end{align*}
where ${\rm ne}_{l}(X_u)$ denotes the set of local neighbors of $X_u$. The inequality $v_u > u$ is included to ignore double counting. Note that ${\cal B}$ is nothing but a convenient ordering of edges in ${\cal G}_{\rm local}$, and hence ${\cal B}$ contains the same information as ${\cal G}_{\rm local}$. It should also be mentioned that our results are not influenced by the particular labelling used.  The $k^{\mathrm{th}}$ row is given by $D_{k,\boldsymbol{.}} = e_{i} - e_{j}$ where $(i,j)$ is the $k^{\mathrm{th}}$ element of the local neighbor sequence and $e_{i}, e_{j}$ denote two canonical basis vectors of $\mathbb{R}^p$ where the $1$'s occur at $i^{\mathrm{th}}$ and $j^{\mathrm{th}}$ position, respectively. This way each pair is represented by a row in the difference matrix. We denote by $D^a$ the $m_a \times p$ sub-matrix of $D$ selecting all the rows with $a^{\mathrm{th}}$ entry being 0. In other words, $D^a$ is the difference matrix corresponding to all the local neighbor pairs not involving $X_a$. The number of local neighbors of $X_a$ in $\Gamma(n)\setminus\{a\}$ is $m_a$. It should be noted that throughout our discussion we shall assume that the local neighborhood structure is known to us, meaning that $D$ is known beforehand and it does not depend on the data.

Now we define the neighborhood-fused LASSO estimate $\hat{\theta}^{a,\lambda,\mu}$ of $\theta^{a}.$ 
\begin{align}\label{nfldefn}
\hat{\theta}^{a,\lambda,\mu} \quad &= \quad \mathrm{argmin}_{\theta: \theta_{a}=0}\left(n^{-1}||X_{a}-X\theta||^{2}+\lambda||\theta||_{1}+\mu||D^{a}\theta||_{1}\right),
\end{align}
where $||x||_1$ denotes the $l_1$-norm of $x$. Penalizing both the $l_1$-norms of $\theta$ and $D^a\theta$ implies parsimony, thus ensuring sparsity and local constancy at the same time. This property helps us in variable selection and thereby leads to neighborhood selection. Note that we are referring to both the non-local and the local neighbors. The neighborhood estimate of node $a$ is defined by the nodes corresponding to non-null coefficients when $X_a$ is regressed on the rest of the variables with the fused lasso penalty in~\eqref{nfldefn}, in other words
\begin{align}\label{nbdestdefn}
\hat{\mathrm{ne}}^{\lambda,\mu}_a \quad &= \quad \left\{b\in\Gamma(n)\setminus\{a\}: \hat{\theta}^{a,\lambda,\mu}_b \ne 0\right\},
\end{align}
where $\hat{\theta}^{a,\lambda,\mu}_b$ denotes the $b$-th component of the vector $\hat{\theta}^{a,\lambda,\mu}$. It is clear that the selected neighborhood depends on the value of $\lambda$ and $\mu$ chosen. Large values of $\lambda$ and $\mu$ will give rise to more sparse solutions. Usually the regularization parameters are chosen by some cross validation criteria. But in this paper we will find a data driven approach for selection of regularization parameters that speeds up computation and ensures asymptotic consistency in model selection. Meinshausen and B\"{u}hlmann derived a data driven choice for $\lambda$ in their paper. We will extend their method for simultaneous selection of $\lambda$ and $\mu$ from the data.

\section{A Discussion of Local Constancy}\label{LocConstDefn}
Here we discuss further formalizations of the notion of local constancy. It should be noted again that the definition is linked with the ideas used by Honorio et al.~\cite{Honorio2009}, but our definition is more general in nature. First we introduce the notion of \emph{diagonal excluded matrix product}. Given a matrix $A$, its zero operator is defined as $Z(A) :=\left( \mathbb{I}(A_{ij} = 0)\right)_{i,j}$ where $\mathbb{I}$ is the indicator function. The diagonal excluded product of two matrices $A$ and $B$ can now be defined as, 
$$A\oslash B := Z(A) \circ AB,$$
where $\circ$ denotes the Hadamard product of matrices. Although the name does not clearly show how diagonals are removed from the product, usually if $A$ is taken to be a difference matrix defined above, this will eventually lead to the exclusion of diagonals of the matrix $B$. 

\subsection{A more quantitative perspective}\label{LocConstDefn:Quant}

First observe that the absolute values of the non-zero entries of $D \oslash \Omega$ are of the form $|\omega_{jk} - \omega_{jk'}|$ where $(k,k')$ corresponds to a pair of local neighbors. In fact, we have $\|D \oslash \Omega\|_{1}$ equals the sum of all the absolute values of the differences $\omega_{ik} - \omega_{jk},\,k \notin \{i,j\}$ for local neighborhood pairs $(i,j)$. In other words, if we think of the entires $\omega_{ik}$ as measuring conditional dependence of two nodes, then $\|D \oslash \Omega\|_{1}$ measures in how different the local neighborhood pairs are with respect to conditional dependence. A small value of $\|D \oslash \Omega\|_{1}$ indicates that the local neighbors behave similarly in this respect.

Thus, one way to define the local constancy property is to impose a bound on the norm of the difference of local neighbors. We say a model exhibit $\left(\epsilon, {\cal G_{\rm local}}, l_p\right)$-local constancy if
$$\big\| D\oslash \Omega\big\|_p < \epsilon$$
where $\big\|\cdot\big\|_p$ denotes the $l_p$ norm. For our purpose, the best candidate would be $p = 1$ for certain desirable properties like sparsity and convexity. However, a different choice of norm would lead to a different solution which might be more appropriate for other types of problems. With $p=1$, this definition coincides with Honorio's local penalty criterion.

It should be noted that the $\epsilon$ parameter controls the degree of local constancy so this is closely related to the tuning parameter we are going to use. This serves as a constraint on certain matrix parameters that we are trying to estimate. A small $\epsilon$ would ensure a high level of local constancy where a large $\epsilon$ would do the opposite. 

\subsection{A Bayesian Perspective}\label{LocConstDefn:Bayes}   
Since the locality information is being used here as a prior information, it makes sense to interpret it from a Bayesian perspective. Let $\pi(\Omega)$ denote a probability distribution on the space of positive definite matrices. Then one can say that an $\left(\epsilon, \delta, {\cal G_{\rm local}}, l_p\right)$-local constancy property holds for this model if
$$P\left(\big\| D\oslash \Omega \big\|_p < \epsilon\right) > 1 - \delta.$$
Here we need to assume that $\delta < 1$. The $\epsilon$ parameter controls the degree of local constancy and the $\delta$ parameter is indicative of our prior belief about the local constancy.

One can also think of defining local constancy by replacing $\Omega$ in sections \ref{LocConstDefn:Quant} and \ref{LocConstDefn:Bayes} by the partial correlation matrix  $\Pi$. Even though the definition makes sense, the natural transition from the difference matrix used in \cite{Honorio2009} to our analogue $D^a$ will not be consistent with it. The  difference terms corresponding to partial correlation matrix will be
$$|\pi_{jk} - \pi_{jk'}| = \frac{1}{\sqrt{\omega_{jj}}}\left| \frac{\omega_{jk}}{\sqrt{\omega_{kk}}} - \frac{\omega_{jk'}}{\sqrt{\omega_{k'k'}}}\right|.$$
But, when we regress $X_j$ on the rest of the nodes, then the local difference we aim to penalize is $\frac{1}{\omega_{jj}}|\omega_{jk} - \omega_{jk'}|$. In order to be consistent with our definition of neighborhood fused lasso, we need to replace $D\oslash \Omega$ by $\left(D \Omega_d \right)\oslash \Pi$, where $\Omega_d = \left({\rm diag}(\Omega)\right)^{\frac{1}{2}}$. Assuming that the individual conditional variances are bounded away from zero (assumption A2 in section~\ref{ass:variance}), this rescaling does not affect the local constancy and our penalty is consistent with the definition. Partial correlations have the advantage of all being on the same scale. This might in particular simplify the choice of $\epsilon.$

\section{Optimization method}\label{optimization}
First we briefly review the existing algorithms used for optimization of fused lasso type problems. Broadly speaking, there are two fundamental classes of algorithms used in this type of problems: (a) Solution path algorithms - which finds the entire solution for all values of the regularization parameters and (b) Approximation algorithms - which attempt to solve a large scale optimization given  a fixed set of regularization parameters using first order approximations.

Friedman et al.~\cite{Friedman2007} formulated the path-wise optimization method for the standard fused lasso signal approximation problem where the design matrix $X = I$. The algorithm was two-step and the final solution is obtained by soft-thresholding the total-variation norm penalized estimate obtained in the first step. The basic challenge in applying the coordinate descent algorithm to a fused lasso problem is the non-separability of the total variation penalty unlike usual lasso where the $l_1$-penalty is completely separable. So they used a modified coordinate descent approach where the \emph{descent} step was followed by an additional fusion and smoothing step. However, this method works only for the total variation penalty and does not extend to fused lasso regression problems with generalized fusion penalty like our situation. Hoefling~\cite{Hoefling2010} proposed a path-wise optimization algorithm for generalized fusion penalties in fused lasso signal approximation problem. This algorithm uses the fact that when varying the penalty parameter, the corresponding sets of fused coefficients do only change at finitely many values of the penalty parameter. Tibshirani and Taylor~\cite{Tibshirani2011} proposed another path algorithm for the fused lasso regression problem with a generalized fused lasso penalty by solving the dual optimization problem. However, the path algorithm they devised can be applied only when the design matrix is full rank and hence is not applicable for high dimensional problems. In order to resolve the problem when the matrix is not full rank, they proposed to add an infinitesimal perturbation $\epsilon \| \beta\|^2$. However, this does not solve the problem completely as a small $\epsilon$ leads to ill-conditioning and increasing number of rows in the generalized fused penalty matrix causes inefficient solutions because of an increasing number of dual variables.

Approximation algorithms were developed to find efficient solutions to general fused lasso problem with a fixed set of penalty parameters, regardless of its rank and they usually adapt themselves to high dimensional problems quite easily. Usually these approximation algorithms are based on first order approximation type methods like gradient descent. Liu et al.~\cite{Liu2010} proposed the efficient fused lasso algorithm (EFLA) that solves standard fused lasso regression problem by replacing the quadratic error term in the optimization function by its first order Taylor expansion at an approximate solution followed by an additional quadratic regularization term. The approximate objective function has a fused lasso signal approximation form and can be solved by applying gradient descent on its dual which is a box-constrained quadratic program. Chen et al.~\cite{Chen2012} proposed the \emph{smoothing proximal gradient} method to solve regression problem with structured penalties that closely resemble our objective function. The basic idea is to approximate the fused penalty term $\| D\beta \|_1$ by a smooth function and devise an iterative scheme for an approximate optimization. They use the smooth function 
$$\tilde{\Omega}(\beta, t) = {\rm max}_{\big\|\alpha\big\|_\infty \leq 1} \left(\alpha^T D\beta - \frac{t}{2}\big\|\alpha\big\|^2_2\right).$$
Convexity and continuous differentiability of this function follows from Nesterov~\cite{Nesterov2005}. One may now proceed using standard first order approximation approach like FISTA~\cite{Beck2009} which works like EFLA. However this process is computationally intensive and not a feasible approach in our case where we need to do a nodewise regression. Another algorithm for tackling a similar problem is known as \emph{Split Bregman} algorithm~\cite{Ye2011}. It was proposed for standard fused lasso regression and later extended to generalized fused lasso regression. The SB algorithm is derived from augmented Lagrangian~\cite{Hestenes1969,Rockafeller1973}, which adds quadratic penalty terms to the penalized objective function and alternatingly solves the primal and dual starting from an initial estimate. This is also computationally intensive unless one has a simple structure constraint on the parameters.

We propose a different algorithm to solve our optimization problem. We show that the neighborhood-fused lasso problem can be reparametrized into a standard lasso problem, thus simplifying the optimization procedure.

The neighborhood-fused LASSO estimate $\hat{\theta}^{a,\lambda,\mu}$ of $\theta^a$ can be written as
\begin{align}
\hat{\theta}^{a,\lambda,\mu} &= \mathrm{argmin}_{\theta: \theta_a = 0} \left(\frac{1}{n}||X_a - X\theta||^2 + \lambda ||\theta||_1 + \mu ||D^a \theta||_1\right)\nonumber\\
&= \mathrm{argmin}_{\theta: \theta_a = 0} \left(\frac{1}{n}||X_a - X\theta||^2 + \lambda \left(||\theta||_1 + \frac{\mu}{\lambda} ||D^a \theta||_1\right)\right)\nonumber\\
&= \mathrm{argmin}_{\theta: \theta_a = 0} \left(\frac{1}{n}||X_a - X\theta||^2 + \lambda \left|\left|\left(\begin{matrix}I\\ \frac{\mu}{\lambda}D^a\end{matrix}\right)\theta\right|\right|_1\right).
\intertext{Letting $G_a = \left(\begin{matrix}I\\ \frac{\mu}{\lambda}D^a\end{matrix}\right)$ and $\omega = G_a\theta$, we define}
\hat{\omega}^{a,\lambda,\mu}&= \mathrm{argmin}_{\omega} \left(\frac{1}{n}||X_a - XG^+_a\omega||^2 + \lambda||\omega||_1\right),\label{omegahat}
\intertext{where $G^+_a$ is the Moore-Penrose inverse of $G_a$. We find $\hat{\theta}^{a,\lambda,\mu}$ from the following relation (see lemma~\ref{optim} for details)}
\hat{\theta}^{a,\lambda,\mu} &= G^+_a \hat{\omega}^{a,\lambda,\mu}.
\end{align}
We note that since $G_a$ has full column rank, $G^+_a$ has full row rank. Thus, the following lemma can be applied here.

\begin{lem}\label{optim}
Let 
$$\tilde{\beta} := \mathrm{argmin}_{\beta\in\mathbb{R}^k} \big\| y - X\beta\big\|^2 + \lambda\big\| G\beta\big\|_1$$ 
and 
$$\tilde{\omega} := \mathrm{argmin}_{\omega} \big\| y - XG^+\omega\big\|^2 + \lambda\big\|\omega\big\|_1.$$ 
Also assume that $G$ is of full column rank. Then $$\tilde{\beta} = G^+\tilde{\omega}.$$
\end{lem}

One interesting observation here is that although we started with the assumption that $\omega = G_a\theta \in \mathcal{C}(G_a)$, where $\mathcal{C}(G_a)$ denotes the column space of $G_a$, we did not optimize over $\mathcal{C}(G_a)$ but did the same over all $\omega$. See the proof of lemma~\ref{optim} for details. The heuristic idea behind this is that we find the minimizer on $\mathcal{C}(G_a)$ by projecting the global minimizer onto $\mathcal{C}(G_a)$ and the projection operator is given by $G_a G^+_a$. The objective function in~\eqref{omegahat} combines the two $\mathit{l}_1$ penalties into a single $\mathit{l}_1$ penalty and thus could be easily solved by any of the standard LASSO algorithms. The parameters $\lambda$ and $\mu$ are chosen according to theorem~\ref{regparam}. We show by several simulations that the proposed method performs better than Meinshausen - B\"{u}hlmann's method or graphical lasso in situations where local constancy holds.

\section{Theoretical Properties of NFL}\label{theory}
\subsection{Model selection consistency and choice of penalty paramteters}
From discussion in section 3, it can be seen that using neighborhood-fused lasso leads to efficient model selection when applied successively to all the nodes. In this section, we show that our proposed method leads to asymptotically consistent model selection similar to the procedure by Meinshausen and B\"{u}hlmann. The choice of the regularization parameters is crucial in such cases. Moreover, we show that our proposed choice of regularization parameters not only ensures convergence to the ``true" model but the convergence is faster than the Meinhausen-B\"{u}hlmann's method when the underlying model is indeed locally constant. This property is also supported by our simulation results shown in section~\ref{simulations}.

\subsubsection{Assumptions}\label{assump}
In order to prove consistency of our method for Gaussian graphical models, we need to work with the following assumptions. Assumptions [A1]-[A3] and [A5]-[A7] are as in Meinshausen and B\"{u}hlmann's (See \cite{Meinshausen2006}, section 2.3). However we need the two additional assumptions [A4] and [A8] to deal with the local constancy.
\begin{description}
\item[{[A1]}]\label{ass:dimensionality} \textbf{Dimensionality} : $\exists \gamma>0$, such that $p(n)=O(n^\gamma)$ as $n\rightarrow \infty$. Note that $\gamma > 1$ is included thus allowing $p \gg n$.
\item[{[A2]}]\label{ass:variance} For all $a\in \Gamma(n)$ and $n\in \mathbb{N}$, $Var(X_a)=1$. There exists $v^2>0$, so that for all $n\in \mathbb{N}$ and $a\in \Gamma(n)$, $Var(X_a | X_{\Gamma(n)\backslash\{a\}})\geq v^2$. This means that all the conditional variances are bounded away from 0.
\item[{[A3]}]\label{ass:sparsity} \textbf{Sparsity} : There exists some $0\leq\kappa<1$ so that $\mathrm{max}_{a\in \Gamma(n)}|\mathrm{ne}_a|= O(n^\kappa)$ for $n\rightarrow \infty$.
\item[{[A4]}]\label{ass:lnsparse} \textbf{Local Neighborhood Sparsity}: The number of local neighbors also can grow at polynomial rate of $n$, i.e., $\exists \beta_0\geq 0$ such that the maximum number of local neighbors of a node $\mathrm{max}_{a\in\Gamma(n)}|{\rm ne}_l(a)|=O(n^{\beta_0})$ for $n\rightarrow\infty$.  For convenience we let $K>0$ be such that
    \begin{align*}2 + \mathrm{max}_{a\in\Gamma(n)}|{\rm ne}_l(a)| < Kn^{\beta_0}.\end{align*}
\item[{[A5]}]\label{ass:bounded} \textbf{$l_1$-Boundedness}: There exists some $\vartheta < \infty$ so that for all neighboring nodes $a,b \in \Gamma(n)$ and all $n\in \mathbb{N}$, $\|\theta^{a,ne_{b}\setminus \{a\}}\|_{1} \leq \vartheta$.
\item[{[A6]}]\label{ass:parcor} \textbf{Magnitude of partial correlation} : There exists a constant $\delta > 0$ and some $\xi > \kappa$ with $\kappa$ as in [A3], so that for every $(a,b)\in E$, $|\pi_{ab}|\geq \delta n^{-\frac{1-\xi}{2}+\beta_0}$ where $\pi_{ab}$ denotes the partial correlation of $X_a$ and $X_b$.
\item[{[A7]}]\label{ass:ns} \textbf{Neighborhood Stability} : Define $S_a(b) = \sum_{k\in\mathrm{ne}_a}\mathrm{sgn}(\theta^{a,\mathrm{ne}_a}_k)\theta^{b,\mathrm{ne}_a}_k$. There exists some $\delta_1<1$ so that for all $a,b\in \Gamma(n)$ with $b\notin \mathrm{ne}_a$, $|S_a(b)|<\delta_1$
\item[{[A8]}]\label{lnstable} \textbf{Local Neighborhood Stability}: Let $\mathcal{L}_a:=\{k: (\left(D^a\right)'\mathrm{sgn}(D^a\theta^a))_k \ne 0\}$ and $T_a(b) := \sum_{k\in\mathcal{L}_a} \left[\left(D^a\right)' \mathrm{sgn}(D^a \theta^a)\right]_k \theta^{b,\mathcal{L}_a}_k$. There exists some $\delta_2 < 1$ so that for all $a,b\in\Gamma(n)$ such that $b\notin \mathcal{L}_a$, $|T_a(b)| < \delta_2 \|D^a_{.b}\|_1$, where $D^a_{.b}$ denotes denotes the $b^{\mathrm{th}}$ column of $D^a$.
\end{description}

Similar to Meinshausen and B\"{u}hlmann's interpretation, we can describe an intuitive condition which implies the last two assumptions. Define $$\theta^a(\eta_1,\eta_2) = \mathrm{argmin}_{\theta:\theta_a=0} E(X_a - \sum_{k\in\Gamma(n)}\theta_k X_k)^2 + \eta_1 \|\theta\|_1 + \eta_2 \|D^a\theta^a\|_1.$$ According to the characterization of $\mathrm{ne}_a$ derived from~\eqref{nbddefn}, $\mathrm{ne}_a = \{k\in\Gamma(n):\theta^a_k(0,0)\ne 0\}$. One can think of a two-dimensional perturbation approach in which one tweaks the parameters $\eta_1$ and $\eta_2$ in a way that the perturbed neighborhood $\mathrm{ne}_{a}(\eta_1,\eta_2) = \{k\in\Gamma(n):\theta^a_k(\eta_1,\eta_2)\ne 0\}$ is identical to original neighborhood ${\rm ne}_a(0,0)$. The following proposition shows that the two assumptions of neighborhood stability are fulfilled under this situation. The terms $S_a(b)$ and $T_a(b)$ measure the sub-gradient of the lasso penalty and the neighborhood-fused lasso penalty respectively. Having a small $l_1$ bound on them enforces the stability of the estimated coefficients, and hence stability of neighbors.  See section~\ref{proofs} for proof of the proposition.

\begin{prop}\label{perturb}
If there exists some $\eta_1 >0, \eta_2 >0$ such that $\mathrm{ne}_a(\eta_1,0)=\mathrm{ne}_a(0,\eta_2)=\mathrm{ne}_a(0,0)$. Then, $|S_a(b)|\leq 1$ and $|T_a(b)|\leq \|D^a_{.b}\|_1$. Moreover, $\mathrm{ne}_a(\eta_1,\eta_2) = \mathrm{ne}_a(0,0)$.
\end{prop}

We start the presentation of the theoretical results with a lemma (proven in section~\ref{proofs}) characterizing the minimizer of our objective function in terms of its subdifferential. 

\begin{lem}\label{baselemma}
Given $\theta\in \mathbb{R}^{p(n)}$, let $G(\theta)$ be a $p(n)$ dimensional vector with elements $G_{b}(\theta) = -\frac{2}{n}\langle X_{a}-X\theta, X_{b} \rangle$. Define $${\cal D}^a_b = [(D^a)'\mathrm{sgn}(D^{a}\theta)]_{b}  \qquad \text{and} \qquad \mathcal{L}^a (\theta)= \{b:{\cal D}^a_b\ne 0\}.$$A vector $\hat{\theta}$ is a solution to the fused LASSO problem described above iff

\begin{align*}
&G_{b}(\hat{\theta})= \lambda \mathrm{sgn}(\hat{\theta}_{b})+\mu {\cal D}^a_b &\text{ when } \hat{\theta}_{b}\ne 0, b\in \mathcal{L}^a(\hat{\theta}),\\
&\lambda\mathrm{sgn}(\hat{\theta}_b) - \mu\big\| D^a_{.b}\big\|_1 \leq G_{b}(\hat{\theta}) \leq \lambda\mathrm{sgn}(\theta_b) + \mu\big\| D^a_{.b}\big\|_1 &\text{ when } \theta_{b}\ne 0, b\notin \mathcal{L}^a(\hat{\theta}),\\
&-\lambda + \mu {\cal D}^a_b \leq G_{b}(\hat{\theta}) \leq \lambda + \mu {\cal D}^a_b &\text{ when } \theta_{b}= 0, b\in \mathcal{L}^a(\hat{\theta}),\\
&|G_b(\hat{\theta})| \leq \lambda + \mu \big\| D^a_{.b}\big\|_1 &\text{ otherwise.\hspace*{2.2cm}}
\end{align*}
\end{lem}

This lemma builds the foundation of several of the following results and will be used frequently to prove them. Sign consistency is one of the major properties that a model selection method should exhibit. Before we study the model selection consistency of our estimators, we show, in the following lemma that the neighborhood-fused lasso estimator is sign-consistent.

\begin{lem}\label{signconsist}
Let $\hat{\theta}^{a,\lambda,\mu}$ be defined for all $a$. Under the assumptions [A1]-[A7], it holds for some $c$ that for all $a$, $\mathbb{P}(\mathrm{sgn}(\hat{\theta}^{a,\lambda,\mu}_{b})=\mathrm{sgn}(\theta^{a}_{b})\forall b\in ne_{a})=1-O(\mathrm{exp}(-cn^{\epsilon})).$
\end{lem}

Observe that this lemma, in turn, preserves the asymptotic equality of signs of local neighbors. If $X_b$ and $X_b'$ are local neighbors, then with high probability, when regressing $X_{a}$ on the remaining variables, the coefficients of  $X_{b}$ and $X_{b'}$ will have the same sign. This is a direct consequence of local constancy of the estimated regression coefficients.

Our results show that, just like in Meinshausen and B\"{u}hlmann, a rate slower than $n^{-1/2}$ is necessary for the regularization parameters for consistent model selection in the high dimensional case where the dimension may increase as a polynomial in the sample size. Specifically if $\lambda$ decays at $n^{-\frac{1-\varepsilon}{2}}$ and $\mu$ decays as $n^{-\frac{1-\varepsilon}{2} - \beta_0}$ where $0<\beta_0<\kappa<\varepsilon<\xi$ are as in assumptions [A1]-[A8], the estimated neighborhood is almost surely contained in the true neighborhood. Hence the type-I error probability goes to 0. This is formally stated in the following theorem (see section~\ref{proofs} for a proof).

\begin{thm}\label{type1}
Let assumptions [A1]-[A8] be fulfilled. Let the penalty parameters satisfy $\lambda_n \sim d_1 n^{-\frac{1-\varepsilon}{2}}$ and $\mu_n \sim d_2 n^{-\frac{1-\varepsilon}{2}}$ with some $\beta_0 \geq 0$ and $0<\kappa<\varepsilon<\xi$ and $d_1, d_2>0$. There exists some $c>0$ such that for all $a\in \Gamma(n)$,
\begin{center}
$P(\hat{\mathrm{ne}}^{\lambda,\mu}_a\subseteq \mathrm{ne}_a) = 1 - O(\exp(-cn^\varepsilon)).$
\end{center}
\end{thm}

The assumptions of neighborhood stability and local neighborhood stability are not redundant. The following proposition shows that one can not relax the assumptions A7 and A8.

\begin{prop}\label{type1problem}
If there exists some $a,b \in \Gamma(n)$ with $b\notin \mathrm{ne}_a$ and $|S_a(b)|>1$ and $|T_a(b)| > \big\| D^a_{.b}\big\|_1$, then for $\lambda_n, \mu_n$ as in theorem 4.6, 
$$P\left(\hat{\mathrm{ne}}^{\lambda,\mu}_a \subseteq \mathrm{ne}_a\right) \rightarrow 0\;\;  \text{as }\;n\rightarrow\infty.$$
\end{prop}

On the other hand, with the same sets of assumptions, one can show that the type II probability, i.e., probability of falsely identifying an edge as a potential connection exponentially goes to 0. This has been formally stated and proved in the following theorem (see section~\ref{proofs} for a proof).

\begin{thm}\label{type2}
With all the assumptions of theorem 4.6 and $\lambda,\mu$ as before,
\begin{align*}
P\left(\mathrm{ne}_a \subseteq \hat{\mathrm{ne}}^\lambda_a\right) = 1 - O(\exp(-cn^\varepsilon)).
\end{align*}
\end{thm}
\subsubsection*{On Model Selection in Graphs}
It follows from the discussion so far that consistent estimation of nodes is possible using our approach. However, one of the original goals of our method is to estimate the entire underlying graphical model, which can be accomplished by combining the estimated neighborhoods in some way. Two different methods of combination have been proposed
\begin{align*}
\hat{E}^{\lambda,\mu,\vee} &:= \left\{(a,b): a \in \hat{\mathrm{ne}}^{\lambda,\mu}_b \vee b \in \hat{\mathrm{ne}}^{\lambda,\mu}_a\right\} \quad (\text{union});\\
\hat{E}^{\lambda,\mu,\wedge} &:= \left\{(a,b): a \in \hat{\mathrm{ne}}^{\lambda,\mu}_b \wedge b \in \hat{\mathrm{ne}}^{\lambda,\mu}_a\right\} \quad (\text{intersection}).
\end{align*}
In our simulations, we combined them following the union method. It has been observed that the differences vanish asymptotically when a regular lasso is applied. Although we did not theoretically study this for our case but our experiments indicate that they do not vary much asymptotically.

\subsubsection*{Finite Sample Choices for Penalty Parameter}
Theoretic exploration in asymptotic domain does not cast much light on the choices of regularization parameter in real life, finite sample problem. It is hard to ensure consistency or absolute containment like theorem~\ref{type1} or~\ref{type2}. However, following the idea proposed by Meinshausen and B\"{u}hlmann, one can consider the \emph{connectivity component} of a node, which is defined as the set of nodes which are connected to it through a chain of edges. Generalizing the results from Meinshausen \& B\"{u}hlmann, the following theorem shows (see proof in section~\ref{proofs}) that the estimated connectivity component derived from neighborhood-fused lasso estimate will belong to the true connectivity component with probability $(1 - \alpha)$, for any chosen level of $\alpha\in(0,1)$.

\begin{thm}\label{regparam}
With all the assumptions [A1]-[A8], and the following choices of the penalty parameters,
\begin{align*}
\lambda &= \frac{\hat{\sigma}_a}{\sqrt{n}} \tilde{\Phi}^{-1}\left(\frac{\alpha}{2p(n)^2}\right)\\
\mu &= \frac{\hat{\sigma}_a}{Kn^{\beta_0 + 1/2}}\tilde{\Phi}^{-1}\left(\frac{\alpha}{2p(n)^2}\right)
\end{align*}
we have
\begin{center}
$P\left(\exists a\in \Gamma(n): \hat{C}^{\lambda,\mu}_a \nsubseteq C_a\right) \leq \alpha$
\end{center}
for all $n$. $C_a$ and $\hat{C}^{\lambda,\mu}_a$ are the true and estimated connectivity components of $a$, $K,\beta_0$ are certain constants and $\tilde{\Phi} = 1 - \Phi$.
\end{thm}

The choice of $K$ and $\beta_0$ depends on the rate of growth of the local neighborhood with increasing sample size. In our simulations, we found that for models with constant dimension and increasing sample size, $K = 1$ and $\beta\in\left(0,\frac{1}{2}\right)$ works fine.\\

As shown in the simulations, we get a higher convergence speed as compared to Meinshausen-B\"{u}hlmann's method by applying a neighborhood-fused lasso penalty for nodewise regression when the underlying model exhibit local constancy. This can be proved using the following lemma (proof in section~\ref{proofs}).

\begin{lem}\label{fastconv}
The upper bound of type I error probability in neighborhood fused lasso is smaller than that of the Meinshausen \& B\"{u}hlmann procedure.\footnote{The lemma \ref{fastconv} delves deep into the precise constants of \eqref{unexplained} in the proof of theorem \ref{type1} and shows that one could achieve smaller type 1 error probability with a finite sample size if one uses neighborhood fused lasso instead of usual lasso. }
\end{lem}
Lemma~\ref{fastconv} does not specify how much the maximal false positive probability is reduced. This can be understood after going through its proof. Roughly speaking, the reduction in the upper bound is given by
    \begin{align*}
    \Big[\, 1 - e^{-\frac{1}{\sigma^2_*} \left(\frac{d_1 d_2}{2}(1 - \delta_1)(1 - \delta_2)n^{\beta_0} + \frac{d^2_2}{4}(1 - \delta_2)^2 n^{2\beta_0}\right) n^\epsilon}\;\Big] \cdot e^{-\frac{1}{\sigma^2_*}\left(\frac{d^2_1}{4}(1-\delta_1)^2 n^{\epsilon}\right)},
    \end{align*}
where $\sigma^2_* = E(X^2_{a,i} V^2_{a,i})$, and $\delta_1$, $\delta_2$ are as in the proof of theorem \ref{type1}. Essentially, it is shown in the proof that the neighborhood fused lasso reduces a dominant portion of the type 1 error (given by the second factor) by a fraction (given by the first factor) shown above. Looking at the proof of theorem~\ref{type1}, it is seen that the term $P\left(\left|2n^{-1}\langle X_a, V_b\rangle\right| \geq (1-\delta_1)\lambda + (1-\delta_2)B\mu\right)$ is the principal contributor to the probability of false positives. The corresponding term using usual lasso is 
$P\left(\left|2n^{-1}\langle X_a, V_b\rangle\right| \geq (1-\delta_1)\lambda\right).$
Lemma~\ref{fastconv} makes use of this fact and it also shows that substantial reduction takes place when the local neighborhood grows. So, this method invariably performs better than the Meinshausen-B\"{u}hlmann's method with respect to a minimax criterion (in a sense that is minimizes the maximum probability of false positives), and the improvement is more when the local neighborhood grows faster. It also implies that it can perform as bad as usual lasso regression as the worst case scenario.

\subsection{Compatibility and $l_1$ properties}\label{compatibility}
In our attempt to Gaussian graphical model learning, we adopt the Meinshausen-B\"{u}hlmann approach, i.e., do a componentwise penalized regression. However, one should keep in mind that in the process of doing so, the design matrix (which is random here) changes at every iteration. In previous section we discussed the asymptotic model selection consistency of our estimator. In this section, we shall go beyond model selection and explore conditions under which our neighborhood-fused lasso estimate exhibit nice asymptotic $l_1$ properties. We shall carry out our theoretical analysis under the assumption that the linear regression model holds exactly, with some underlying ``true" $\theta^0$. Most of the theoretical results in this section have been derived in the light of discussions in \cite{Buhlmann2011}.

We start with a quick recapitulation of the notation we are going to use here. If $X = (X_1, X_2, \cdots, X_p) \sim N(0,\Sigma)$, one assumes a node-wise regression model
\begin{align}\label{regmodel}
X_a &= X^a \theta_a + \epsilon_a \text{ for } a = 1,2,\cdots, p,
\end{align}
where $X_a$ denotes the $a$-th component, $X^a$ denotes all the components except $a$ and $\epsilon \sim N(0,\sigma^2 I_n)$ for some $\sigma^2$. Also assume that $\Sigma := \left(\left(\sigma_{ij}\right)\right)_{i,j=1,\ldots,p}$ and $\Omega := \Sigma^{-1} = \left(\left(\sigma^{ij}\right)\right)_{i,j=1,\ldots,p}$ is the precision matrix. If $E$ denotes the edge set in the conditional independence graph then $(i,j) \notin E \Leftrightarrow \sigma^{ij} = \sigma^{ji} = 0$. The design matrix $X^a$ is obtained by deleting the $a$-th column of the data matrix.

In the high dimensional set up, where one generally has lesser number of samples that model dimension $(n < p)$, we assume an inherent sparsity in the true $\theta^a$. This sparsity is also ensured if we assume that the underlying conditional independence graph is sparse. Let us assume
\begin{align*}
S^0_a &= \{j: \theta_{a,j} \ne 0\}
\end{align*}
and $s^0_a = \mathrm{card}(S^0_a)$. Since $S^0_a$ is not known, one needs a regularization penalty. Meinshausen and B\"{u}hlmann chose the $l_1$ penalty, i.e., the traditional lasso and got the estimate
\begin{align*}
\hat{\theta}^\lambda_a &= \mathrm{argmin}_{\theta_a} \left[\frac{1}{n}\big\| X_a - X^a \theta_a\big\|^2 + \lambda\big\|\theta_a\big\|_1\right].
\end{align*}
In our situation, we have added another penalty term $\big\| D^a\theta_a\big\|_1$ along with the lasso penalty term. Hence, our estimator is given by
\begin{align*}
\hat{\theta}^{\lambda,\mu}_a &= \mathrm{argmin}_{\theta_a} \left[\frac{1}{n}\big\| X_a - X^a \theta_a\big\|^2 + \lambda\big\|\theta_a\big\|_1 + \mu\big\| D^a\theta_a\big\|_1\right].
\end{align*}
Note that this new definition of our estimator is exactly same as what was defined in~\ref{nfldefn}.

\subsection*{The Compatibility Condition for NFLasso}
As mentioned in the earlier section, we shall develop our theory based on the assumption of a linear truth. We try to provide an upper bound on the prediction error. The following lemma forms the basis of our derivation.

\begin{lem}{\textbf{The basic inequality}}\label{basicineq}
\begin{align*}
\frac{1}{n}\big\| X^a (\hat{\theta}^{\lambda,\mu}_a - \theta^0_a)\big\|^2 + &\lambda\big\|\hat{\theta}^{\lambda,\mu}_a\big\|_1  + \mu\big\| D^a\hat{\theta}^{\lambda,\mu}_a\big\|_1 \\
&\leq \frac{2}{n}\epsilon'_a X^a(\hat{\theta}^{\lambda,\mu}_a - \theta^0_a) + \lambda\big\|\theta^0_a\big\|_1 + \mu\big\| D^a\theta^0_a\big\|_1.
\end{align*}
\end{lem}

It should be noted that the number of non zero elements in the column of the matrix $D^a$ denotes the number of local neighbors that particular node has (except node $a$). Let us assume that the number of local neighbors is $O(n^{\beta_0})$. Fixing $n$, let us also assume that the number of local neighbors is bounded by $B$. Then, it follows from lemma~\ref{basicineq} using $\left\| \big\| x \big\|_1 - \big\| y \big\|_1\right \|_1 \leq \left\| x - y \right\|_1$ that
\begin{align*}
\frac{1}{n}\big\| X^a (\hat{\theta}^{\lambda,\mu}_a - \theta^0_a)\big\|^2 &\leq \frac{2}{n}\epsilon'_a X^a(\hat{\theta}^{\lambda,\mu}_a - \theta^0_a) + (\lambda + B \mu) \big\|(\hat{\theta}^{\lambda,\mu}_a - \theta^0_a)\big\|_1.
\end{align*}
The basic objective of using a penalty parameter is to overrule the empirical process term $\frac{2}{n}\epsilon'_a X^a(\hat{\theta}^{\lambda,\mu}_a - \theta^0_a)$. It can be easily seen that
\begin{align*}
\left|\frac{2}{n}\epsilon'_a X^a(\hat{\theta}^{\lambda,\mu}_a - \theta^0_a)\right| &\leq \left( \frac{2}{n} \mathrm{max}_{1\leq j\leq p} \left|\epsilon'_a X^a_j\right|\right)\big\| \hat{\theta}^{\lambda,\mu}_a - \theta^0_a \big\|_1.
\end{align*}
Our goal is to choose a $\lambda$ and a $\mu$ such that the probability that the empirical process term in the right hand side exceeds $\lambda + B\mu$ is small, so that with high probability the right hand side could be dominated by $(\lambda +B\mu) \big\| \hat{\theta}^{\lambda,\mu}_a - \theta^0_a \big\|_1$. To that effect, we define
\begin{align*}
\Lambda_a := \left\{\mathrm{max}_{\stackrel{1\leq j\leq p}{j\ne a}} \frac{2}{n} \left|\epsilon'_a X^a_j\right| \leq \lambda_0 + B\mu_0\right\}.
\end{align*}
Our objective is to show that for some particular choice of $\lambda_0$ and $\mu_0$, $\Lambda_a$ has high probability. Now one should keep in mind that both $\epsilon_a$ and $X^a_j$ are random here. One can make the valid assumption of their individual Gaussian law and independence. We formalize these notions in the following proposition.
\begin{prop}
Under the assumption of linear truth for the Gaussian graphical model, i.e., $X_a = X^a \theta^0_a + \epsilon_a \text{ for } a = 1,2,\cdots, p$, we have 
$$\epsilon_{a,i} \stackrel{\mathrm{i.i.d.}}{\sim} N\left(0, \sigma_{aa} - \Sigma_{ab}\Sigma^{-1}_{bb}\Sigma'_{ab}\right)\;\; \text{where }\;
\Sigma = \left(
\begin{matrix}
\sigma_{aa} & \Sigma_{ab}\\
\Sigma'_{ab} & \Sigma_{bb}
\end{matrix}
\right).$$
\end{prop}
The proof of this proposition is straightforward and hence skipped. The following results show that for suitable choice of $\lambda_0$ and $\mu_0$, $P(\Lambda_a)$ is high.
\begin{lem}\label{highprob}
Under the assumption that $\sigma_i = 1 \quad \forall i = 1, 2, \cdots, p$,
\begin{align*}
P(\Lambda_a) & \geq 1 - 2 \exp\left[\log p - n\left(\frac{\lambda_0}{2} + 1 - \sqrt{\lambda_0 + B \mu_0 + 1}\right)\right],
\intertext{In particular, with $\lambda_0 = \frac{2(t + \log p)}{n} > 0$ and $\mu_0 = \frac{2}{B}\sqrt{\frac{2}{n}(t + \log p)}$, then}
P(\Lambda_a) &\geq 1 - 2 e^{-t}.
\end{align*}
\end{lem}
\begin{cor}\label{highprobcor}
Assume that $\sigma_i = 1 \quad \forall i = 1, 2, \cdots, p$ and that $p = O(n^\gamma)$. Let the regularization parameters be
\begin{align*}
\lambda &= \frac{2\left(t^2 + \log p\right)}{n},\quad \mu = \frac{1}{B} \sqrt{\frac{8(t^2 + \log p)}{n}}.
\end{align*}
Then the following is true
\begin{align*}
& P\left(\frac{1}{n}\big\| X^a (\hat{\theta}^{\lambda,\mu}_a - \theta^0_a)\big\|^2 \leq 2(\lambda + B\mu)\big\| \theta^0_a \big\|_1 + \mu B\big\|\hat{\theta}^{\lambda,\mu}_a\big\|_1 \right) \geq 1 - e^{-t^2}.
\end{align*}
\end{cor}
The following lemma provides an upper bound on the estimation error.
\begin{lem}\label{l2ineq}
Assume that $\sigma_i = 1 \quad \forall i = 1, 2, \cdots, p$ and that $p = O(n^\gamma)$. Let the regularization parameters be
\begin{align*}
\lambda = \frac{2\left(t^2 + \log p\right)}{n}, \qquad
& \mu = \frac{1}{B} \sqrt{\frac{8(t^2 + \log p)}{n}}.
\end{align*}
Also let $\delta_{\mathrm{min}}$ be the smallest non-zero singular value of $X^a$. Then we have
\begin{align*}
P\Big[ \frac{1}{n}\| &X^a (\hat{\theta}^{\lambda,\mu}_a - \theta^0_a)\|^2 \\
& \leq 2\left(\lambda + B\mu\left(1 + \frac{\sqrt{p}}{2}\right)\right)\| \theta^0_a \|_1 + \frac{npB\mu(\lambda + B\mu)}{\delta_{\mathrm{min}}}\Big] \geq 1 - e^{-t^2}.
\end{align*}
\end{lem}

\subsubsection*{Oracle Inequalities}
We now try to derive some oracle inequalities which will provide $l_1$ bound for our neighborhood-fused lasso estimate. Following B\"{u}hlmann's notation, let us write, for an index set $S \subset \{1,2,\cdots, p\}$,
\begin{align*}
\theta_{a,j,S} &:= \theta_{a,j}\mathbf{1}\{j\in S\},\\
\theta_{a,j,S^c} &:= \theta_{a,j}\mathbf{1}\{j\notin S\}.
\end{align*}
We need some more notation. Split the matrix $D^a$ as follows:
\begin{align*}
D^a &= \left[
\begin{matrix}
D^a_{S,S} & 0\\
D^a_{S,0} & D^a_{0,S^c}\\
0 & D^a_{S^c,S^c}
\end{matrix}\right],
\end{align*}
where $D^a_{S,S}$ consists of all rows such that both the non zero terms belong to $S$.
$D^a_{S,0}$ consists of all rows and columns such that exactly one of the non-zero term in that row belongs to $S$.
$D^a_{0,S^c}$ consists of all rows and columns such that exactly one of the non-zero term in that row belongs to $S^c$.
$D^a_{S^c,S^c}$ consists of all rows such that both the non-zero terms belong to $S$.

\begin{lem}\label{oracle1}
On $\Lambda_a$, if we choose $\lambda \geq 2\lambda_0$ and $\mu \geq 2\mu_0$, we have
\begin{align*}
\frac{2}{n}\big\| X^a \left(\hat{\theta}^{\lambda,\mu}_a - \theta^0_a\right)\big\|^2 + (\lambda - 3B\mu) \big\|\hat{\theta}^{\lambda,\mu}_{a,S^c_0}\big\|_1 \quad &\leq\quad (3\lambda+5B\mu) \big\|\hat{\theta}^{\lambda,\mu}_{a,S_0} - \theta^0_{a,S_0}\big\|_1.
\end{align*}
\end{lem}
It can be easily verified that if $\lambda \geq \left(3 + \frac{14}{\Delta}\right)B\mu$ for some $\Delta>0$, we have
\begin{align*}
\lambda - 3B\mu &\geq \frac{1}{3+\Delta} (3\lambda + 5B\mu).
\end{align*}
This helps us to simplify the consequences lemma~\ref{oracle1}, which implies that
\begin{align*}
(\lambda - 3B\mu) \big\|\hat{\theta}^{\lambda,\mu}_{a,S^c_0}\big\|_1 \quad &\leq\quad (3\lambda+5B\mu) \big\|\hat{\theta}^{\lambda,\mu}_{a,S_0} - \theta^0_{a,S_0}\big\|_1.
\end{align*}
Combining with the aforementioned condition, we get
\begin{align*}
\frac{3\lambda + 5B\mu}{3+\Delta}\|\hat{\theta}^{\lambda,\mu}_{a,S^c_0}\|_1 \;\;\leq\;\;  &(\lambda - 3B\mu) \|\hat{\theta}^{\lambda,\mu}_{a,S^c_0}\|_1 \;\; \leq\;\;(3\lambda+5B\mu) \|\hat{\theta}^{\lambda,\mu}_{a,S_0} - \theta^0_{a,S_0}\|_1,\\
\intertext{or,}
\big\|\hat{\theta}^{\lambda,\mu}_{a,S^c_0}\big\|_1 \;\;&\leq\quad (3+\Delta)\big\|\hat{\theta}^{\lambda,\mu}_{a,S_0} - \theta^0_{a,S_0}\big\|_1.
\end{align*}
A standard way to relate the $l_1$ penalty $\big\|\hat{\theta}^{\lambda,\mu}_{a,S_0} - \theta^0_{a,S_0}\big\|_1$ to an $l_2$ penalty is to use the Cauchy-Schwarz inequality
\begin{align*}
\big\|\hat{\theta}^{\lambda,\mu}_{a,S_0} - \theta^0_{a,S_0}\big\|_1 &\leq \sqrt{s_0} \big\|\hat{\theta}^{\lambda,\mu}_{a,S_0} - \theta^0_{a,S_0}\big\|,
\intertext{and subsequently relate it to the $l_2$ penalty on the left hand side}
\frac{2}{n}\big\| X^a\left(\hat{\theta}^{\lambda,\mu}_a - \theta^0_a\right)\big\|^2 &= \frac{2}{n} \left(\hat{\theta}^{\lambda,\mu}_a - \theta^0_a\right)' X^{a'}X^a \left(\hat{\theta}^{\lambda,\mu}_a - \theta^0_a\right)
\intertext{in the following manner}
\big\|\hat{\theta}^{\lambda,\mu}_{a,S_0} - \theta^0_{a,S_0}\big\|^2 &\leq \frac{ \left(\hat{\theta}^{\lambda,\mu}_a - \theta^0_a\right)' X^{a'}X^a \left(\hat{\theta}^{\lambda,\mu}_a - \theta^0_a\right)}{\phi^2_{0,a}},
\end{align*}
where $\phi_{0,a} > 0$ is some constant. However, $\hat{\theta}^{\lambda,\mu}_a$ being random, this condition can not hold unanimously for all $\theta\in\mathbb{R}^p$. B\"{u}hlmann provided a \emph{compatibility condition} so that this is true. In our situation, we found a similar condition like them to carry out further analysis. It should be noted that our compatibility condition is weaker than the compatibility condition B\"{u}hlmann provided. However, we need to work under the assumption that $\lambda \geq \left(3 + \frac{14}{\Delta}\right)$. The definition of this \emph{compatibility condition} is provided below.
\begin{defn}\label{compat} \hspace*{-0.2cm}(Compatibility Condition)
We say that a collection of nodes $S_0$  satsifies the $\Delta$-compatibility condition if for all $\theta$ satisfying
\begin{align*}
\big\|\theta_{S^c_0}\big\|_1 &\leq (3 + \Delta)\big\|\theta_{S_0} \big\|_1
\end{align*}
we have
\begin{align*}
\big\|\theta_{S_0}\big\|^2_1 &\leq \frac{s_0\left(\theta' X^{a'}X^a \theta\right)}{n\phi^2_{0,a}}.
\end{align*}
\end{defn}
Following Bickel et al.~\cite{Bickel2009}, we shall refer to $\phi^2_{0,a}$ as the \emph{restricted eigenvalue}. Here we provide a detailed explanation of this phenomenon. Observe that the compatibility condition could be re-written as
\begin{align*}
&\phi^2_{0,a} \leq \mathrm{inf}_{\theta: \big\|\theta_{S^c_0}\big\|_1 \leq (3+\Delta)\big\|\theta_{S_0} \big\|_1} \frac{s_0\left(\frac{1}{n}\theta' X^{a'}X^a \theta\right)}{\big\|\theta_{S_0}\big\|^2_1},
\intertext{so that $\phi_{0,a}$ could be taken as the square root of the infimum assumed by the right hand side. Observing further that}
&\mathrm{inf}_{\theta: \big\|\theta_{S^c_0}\big\|_1 \leq (3+\Delta)\big\|\theta_{S_0} \big\|_1} \frac{s_0\left(\frac{1}{n}\theta' X^{a'}X^a \theta\right)}{\big\|\theta_{S_0}\big\|^2_1} \geq \mathrm{inf}_{\theta} \frac{s_0\left(\frac{1}{n}\theta' X^{a'}X^a \theta\right)}{\big\|\theta_{S_0}\big\|^2_1}\\
&\geq \mathrm{inf}_{\theta} \frac{s_0\left(\frac{1}{n}\theta' X^{a'}X^a \theta\right)}{s_0\big\|\theta_{S_0}\big\|^2}
\geq \mathrm{inf}_{\theta} \frac{\left(\frac{1}{n}\theta' X^{a'}X^a \theta\right)}{\big\|\theta\big\|^2}
= \lambda_{\mathrm{min}}\left(\frac{1}{n}X^{a'}X^a\right),
\end{align*}
$\phi^2_{0,a}$ can be assumed to be the minimum eigenvalue over the restricted set $\{\theta: \big\|\theta_{S^c_0}\big\|_1 \leq (3+\Delta)\big\|\theta_{S_0} \big\|_1\}$. With this compatibility condition imposed, we derive the following oracle inequality
\begin{thm}\label{oracle2}
Assume that the compatibility condition (from definition~\ref{compat}) holds for some $\Delta > 0$. Then on $\Lambda_a$, for $\lambda \geq 2\lambda_0$, $\mu\geq 2\mu_0$ and $\lambda \geq \left(3 + \frac{14}{\Delta}\right)B\mu$, we have
\begin{align*}
\frac{1}{n}\big\| X^a \left(\hat{\theta}^{\lambda,\mu}_a - \theta^0_a\right)\big\|^2 + (\lambda - 3B\mu) \big\|\hat{\theta}^{\lambda,\mu}_a - \theta^0_a\big\|_1 \quad&\leq\quad \frac{s_0(2\lambda+B\mu)^2}{\phi^2_{0,a}}.
\end{align*}
\end{thm}
Combining theorem~\ref{oracle2} and lemma~\ref{highprob}, we get,
\begin{thm}\label{finalthm}
Let 
\begin{align*}
&\lambda = \frac{4 (t + \log p)}{n}, \quad
\mu = \frac{4}{B}\sqrt{\frac{2}{n}(t + \log p)} \quad
\text{and} \quad
t \geq 2n\left(3 + \frac{14}{\Delta}\right)^2,
\end{align*}
and assume that $\sigma_i = 1,\; i = 1,2,\cdots,p$. Then with probability $1 - e^{-t}$,
\begin{align*}
\frac{1}{n}\big\| X^a \left(\hat{\theta}^{\lambda,\mu}_a - \theta^0_a\right)\big\|^2 + (\lambda - 3B\mu) \big\|\hat{\theta}^{\lambda,\mu}_a - \theta^0_a\big\|_1 \quad&\leq\quad \frac{s_0(2\lambda+B\mu)^2}{\phi^2_{0,a}}.
\end{align*}
\end{thm}
Theorem~\ref{finalthm} provides a unified way to deal with both the squared estimation error and the absolute error of the estimated parameters in neighborhood fused lasso regression. It can be seen that for increasing $n$, with probability increasing to 1, the absolute difference of our NFL estimator from underlying truth is bounded by
$$\frac{s_0 (2\lambda + B\mu)^2}{\phi^2_{0,a} (\lambda - 3B\mu)}.$$ 
For large values of $n$, this is approximately equal to
$$\frac{s_0\left(\frac{8t}{n} + 4\sqrt{\frac{2t}{n}}\right)^2}{\phi^2_{0,a}\left(\frac{4t}{n} - 12\sqrt{\frac{2t}{n}}\right)},$$
which is a positive and decreasing function of $\frac{t}{n}$ if $\frac{t}{n} > 18$. This is automatically satisfied for all $\Delta > 0$. So, it can be easily seen, using the assumption $t \geq 2n\left(3 + \frac{14}{\Delta}\right)^2
$, that the above quantity is bounded by 
$$\frac{224 s_0 (\Delta+4)^2(3\Delta+14)}{\phi^2_{0,a}\Delta^2}.$$

\newpage
\section{Simulations}\label{simulations}
\subsection{Simulation 1}\label{sim1}
\begin{figure}[h]
    \centering
  \includegraphics[width=\textwidth]{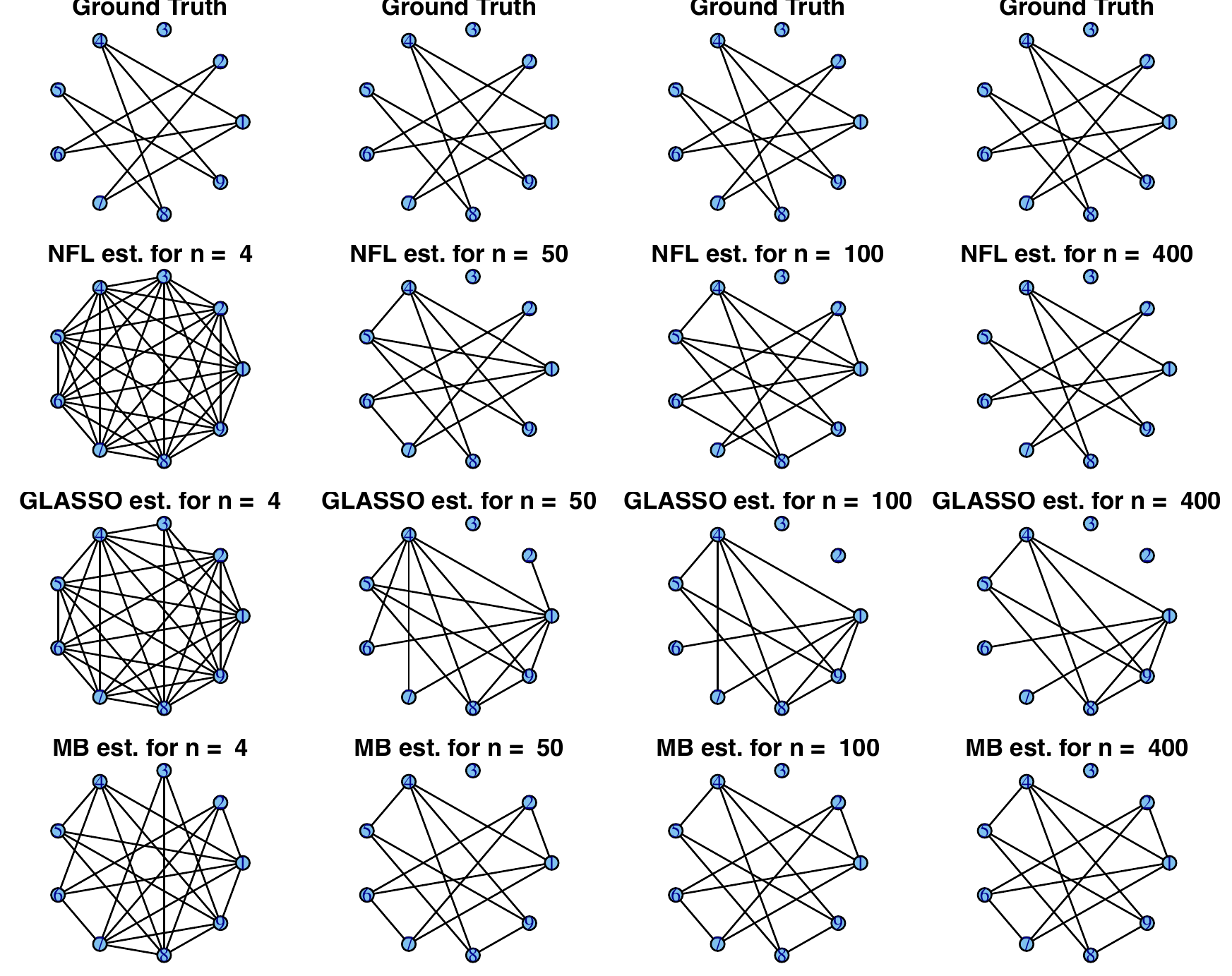}\\
  \caption{Comparison of NFL with GLASSO and Meinshausen-B\"{u}hlmann estimates in section 7.1}\label{nflvscdd}
\end{figure}

In this simulation we repeat the exact scenario presented in Honorio's first simulation setting. The Gaussian graphical model consists of 9 variables as shown in figure \ref{nflvscdd}. It deals with both local and non-local interactions. Our method is compared to Meinshausen-B\"{u}hlmann's method and graphical lasso. We run our simulation for 4 different sample sizes $n = 4, 50, 100 $ \& $400$. For each sample size, we run the simulation 50 times and estimate the neighborhood for each iteration. We construct a weighted graph with edge weight corresponding to the frequency of its occurrence in all of those 50 iterations.

\subsection{Simulation 2}\label{sim2}
\begin{figure}[h]
    \centering
  \includegraphics[width=\textwidth]{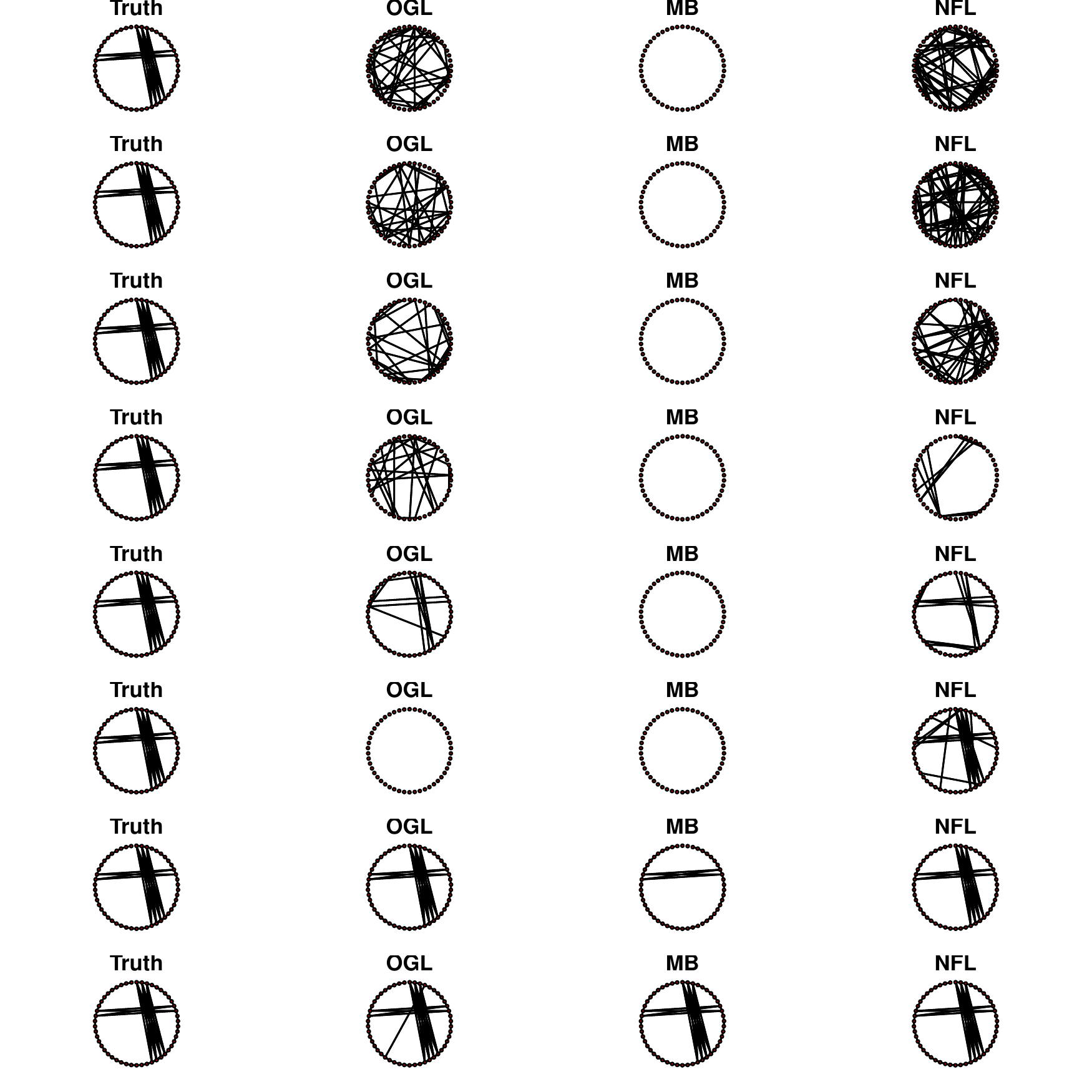}\\
  \caption{Comparison of NFL with GLASSO and Meinshausen-B\"{u}hlmann estimate in section 7.2, sample sizes from top to bottom are 10, 25, 50, 100, 500, 1000, 5000, 10000}\label{bigmodelcomp}
\end{figure}

In this simulation study we take a 50 dimensional normal random vector with zero mean. The diagonals of the precision matrix are all $1$ and all the nonzero off-diagonal entries are $0.2$. The conditional (in)dependence graph of this vector consists of both spatial (local) and non-spatial neighbors where the local neighborhood structure is linear (one dimensional lattice). There are two groups of distant neighbors. We generate $n$ i.i.d. samples from the corresponding normal distribution. We run the simulation for $n = 10, 25, 50, 100, 500, 1000$. We try to reconstruct the conditional (in)dependence graph from the data using Graphical LASSO (we use an oracle version of graphical lasso where the choice of penalty parameter is contingent on the actual number of edges in the true model), Meinshausen - B\"{u}hlmann's coordinate-wise LASSO and our coordinate-wise generalized Fused LASSO approach. For each $n$, we run the simulation 50 times and calculate the number of correctly identified edges and that of falsely identified edges for each iteration. The mean and standard deviations are shown in the table \ref{bigmodeltable1} and table \ref{bigmodeltable2}. Figure \ref{bigmodelcomp} shows the relative performance of the competing methods for different sample sizes. Sample size increases from top to bottom. In the figure we include comparison for two additional sample values. They are $n = 5000$ and $n = 10000$ respectively.

\begin{table}[h]

\hspace{-1in}
\begin{tabular}{|c|c|c|c|c|c|c|c|}
\toprule\toprule
Method & Parameters & \multicolumn{2}{|c|}{$n=10$} & \multicolumn{2}{|c|}{$n=25$} & \multicolumn{2}{|c|}{$n=50$}\\
\midrule
& & fp & tp & fp & tp & fp & tp \\ \hline
GL & $\rho=0.00$ & NA & NA & NA & NA & NA & NA\\ \cline{2-8}
   & $\rho=0.05$ & 239.78(8.94) & 24.98(4.13) & 352.79(9.89) & 40.33(3.86) & 364.65(14.77) & 52.08(3.79)\\ \cline{2-8}
   & $\rho=0.10$ & 217.71(8.73) & 23.20(3.83) & 273.30(11.22) & 36.10(3.57) & 248.21(11.54) & 45.58(3.52)\\ \cline{2-8}
   & $\rho=0.15$ & 195.75(9.85) & 21.78(3.68) & 210.74(11.01) & 31.92(3.88) & 165.44(9.21) & 38.640(3.70)\\ \cline{2-8}
   & $\rho=0.20$ & 175.32(10.02) & 20.60(3.25) & 161.56(10.49) & 27.88(3.45) &  104.28(8.24) & 31.64(3.75)\\ \cline{2-8}
   & $\rho=0.25$ & 156.95(11.18) & 18.86(3.28) & 121.98(10.66) & 23.42(3.84) & 59.92(6.64) & 24.740(4.13)\\ \cline{2-8}
   & $\rho=0.30$ & 138.76(10.79) & 17.32(3.42) & 89.40(10.55) & 19.34(3.51) & 33.08(5.67) & 17.53(3.15) \\ \midrule\midrule

MB & & 0(0) & 0(0) & 0(0) & 0(0) & 0(0) & 0(0) \\ \midrule\midrule
NFL & & 51.62(5.74) & 7.54(2.69) & 37.32(4.84) & 10.12(2.69) & 47.20(6.05) & 20.18(4.02) \\
\bottomrule\bottomrule
\end{tabular}
\caption{Comparison of False Positives and True Positives}\label{bigmodeltable1}
\end{table}

\begin{table}[h]
\centering
\begin{tabular}{|c|c|c|c|c|c|c|c|}
\toprule\toprule
Method & Parameters & \multicolumn{2}{|c|}{$n=100$} & \multicolumn{2}{|c|}{$n=500$} & \multicolumn{2}{|c|}{$n=1000$}\\
\midrule
& & fp & tp & fp & tp & fp & tp \\ \hline
GL & $\rho=0.00$ & 579.72(22.16) & 64.06(3.03) & 580.64(20.99) & 69(0) & 584.7(24.48) & 69(0)\\ \cline{2-8}
   & $\rho=0.05$ & 336.84(15.68) & 60.34(2.79) & 166.10(8.39) & 68.24(0.85) & 83.16(7.06) & 68.9(0.36)\\ \cline{2-8}
   & $\rho=0.10$ & 187.14(11.43) & 52.46(2.97) & 23.90(4.49) & 61.26(1.66) & 2.7(1.76) & 62.74(1.64)\\ \cline{2-8}
   & $\rho=0.15$ & 94.48(9.24) & 42.50(3.59) & 1.5(1.04) & 49.62(1.71) & 0.02(0.14) & 50.28(1.84)\\ \cline{2-8}
   & $\rho=0.20$ & 38.80(6.33) & 32.48(3.25) & 0.04(0.20) & 33.62(2.92) & 0(0) & 35.14(3.09)\\ \cline{2-8}
   & $\rho=0.25$ & 14.30(4.45) & 21.76(3.32) & 0(0) & 15.12(3.99) & 0(0) & 12.86(2.67)\\ \cline{2-8}
   & $\rho=0.30$ & 3.96(2.08) & 13.32(2.87) & 0(0) & 3.6(1.96) & 0(0) & 1.12(0.87)\\ \midrule\midrule

MB & & 0(0) & 0(0) & 0(0) & 0(0) & 0(0) & 0.95(0.92)\\ \midrule\midrule
NFL & & 99.92(7.08) & 41.10(3.73) & 2.44(1.40) & 49.86(2.42) & 0.02(0.14) & 49.7(2.25)\\
\bottomrule\bottomrule
\end{tabular}
\caption{Comparison of False Positives and True Positives}\label{bigmodeltable2}
\end{table}

It is quite evident from the two simulation that our method converges to the true model faster than Meinshausen-B\"{u}hlmann's method. It is also shown theoretically in lemma~\ref{fastconv}.

\newpage
\section{Discussion and Future Works}
In this paper, we have successfully extended the Meinshausen-B\"{u}hlmann approach of model selection in Gaussian graphical models where the assumption of local constancy holds. We also provided a rigorous and generalized definition of local constancy and used it to propose neighborhood-fused lasso (NFL), an algorithm to solve the problem by penalizing the differences of local neighbors given by a pre-determined graph ${\cal G}_{\rm local}$. We have used a generalized version of fused lasso in order to accomplish our objective. Our algorithm reduces the fused lasso problem into a regular lasso problem for a given set of regularizing parameters and thereby fast solutions are readily available. We were able to provide data dependent choices for the tuning parameters in order to establish asymptotic consistency in model selection. We substantiated our theoretical discussion on asymptotic model selection consistency with simulations that furnished desired results. We also proved, both theoretically and experimentally, that incorporating local constancy ensures faster convergence to the underlying truth, meaning that similar accuracy is achieved with smaller sample size. This phenomenon is somewhat similar to what Buhl~\cite{Buhl1993} and Uhler~\cite{Uhler2012} observed while investigating the existence of MLE's in Gaussian models with certain geometric structures. They were able to prove the existence of MLE of a higher dimensional model with relatively smaller sample size. Our findings are somewhat reminiscent of these results. We also discussed about the compatibility issue while doing regression with local constancy and were able to derive sufficient conditions under which $l_1$ boundedness of estimated parameters is ensured. On top of that we also provided an upper bound for the quadratic prediction error.\\\\
There are many extensions possible for future research. Firstly, instead of using the pre-defined local graph ${\cal G}_{\rm local}$, one can think of alternative ways to grow or shrink the local neighborhood adaptively. Secondly, as discussed in section \ref{LocConstDefn}, one can think of devising an algorithm that penalizes the differences of partial correlations instead of the elements of the precision matrix. Thirdly, (because of the compatibility issue) regular lasso or fused lasso might fail to perform up to the mark if the model is ill-conditioned or the compatibility assumptions do not hold. One can think of using adaptive lasso or modify it accordingly in order to incorporate the local constancy property. Some preliminary simulations in this direction look promising. Thirdly, a Bayesian approach could be pursued where one can start with some prior distribution on the inverse covariance matrix respecting the local constancy property and use the Bayesian definition of local constancy in this paper (see~\ref{LocConstDefn:Bayes}) to come up with a novel alternative method.\\
 
\newpage
{\sc \large Appendix.}

\section{Proofs of Theoretical Results}\label{proofs}

\begin{proof}[{\rm \textbf{Proof of lemma~\ref{optim}}}]
Let $G^+ \tilde{\omega} = \beta_0$. We will show that $\beta_0 = \tilde{\beta}$. We know from definition of $\tilde{\omega}$ that $\forall \omega$
\begin{align*}
& \big\| y - XG^+\tilde{\omega}\big\|^2 + \lambda\big\|\tilde{\omega}\big\|_1 \leq \big\| y - XG^+\omega\big\|^2 + \lambda\big\|\omega\big\|_1.
\intertext{Since the objective function is convex, the minimizer in $\mathcal{C}(G)$ is obtained by projecting the grand minimizer onto $\mathcal{C}(G)$. Hence,}
& \hat{\omega} := \mathrm{argmin}_{\omega\in\mathcal{C}(G)} \big\| y - XG^+\omega\big\|^2 + \lambda\big\|\omega\big\|_1 = GG^+\tilde{\omega}.
\intertext{Therefore, for any $\omega\in\mathcal{C}(G)$, we have}
& \big\| y - XG^+\hat{\omega}\big\|^2 + \lambda\big\|\hat{\omega}\big\|_1 \leq \big\| y - XG^+\omega\big\|^2 + \lambda\big\|\omega\big\|_1\\
\Rightarrow & \big\| y - XG^+ G G^+ \tilde{\omega}\big\|^2 + \lambda\big\| G G^+ \tilde{\omega}\big\|_1 \leq \big\| y - XG^+\omega\big\|^2 + \lambda\big\|\omega\big\|_1\\
\Rightarrow & \big\| y - X\beta_0\big\|^2 + \lambda\big\| G \beta_0\big\|_1 \leq \big\| y - XG^+G\theta\big\|^2 + \lambda\big\| G\theta\big\|_1 \quad \forall \theta.
\intertext{Since $G$ is full column rank, $G^+ G = I$. Hence,}
& \big\| y - X\beta_0\big\|^2 + \lambda\big\| G \beta_0\big\|_1 \leq \big\| y - X\theta\big\|^2 + \lambda\big\| G\theta\big\|_1 \quad \forall \theta.
\intertext{Therefore we get that}
& \beta_0 = \mathrm{argmin}_{\beta\in\mathcal{C}(G^+) = \mathbb{R}^k} \big\| y - X\beta\big\|^2 + \lambda\big\| G\beta\big\|_1.
\end{align*}
\end{proof}

\begin{proof}[{\rm \textbf{Proof of proposition~\ref{perturb}}}]
The proof can be found in \ref{suppA}.
\end{proof}

\begin{proof}[{\rm \textbf{Proof of lemma~\ref{baselemma}}}]
The subdifferential of $\frac{1}{n}\big\| X_{a}-X\theta\big\|^{2}_{2}+\lambda \big\|\theta\big\|_{1}+\mu \big\| D^{a}\theta\big\|_{1}$ is given by $\{G(\theta)+\lambda e_{1} + \mu e_{2}: e_{1}\in S_{1}, e_{2}\in S_{2}\}$ where $S_{1}=\{e\in \mathbb{R}^{p(n)}: e_{b}=\mathrm{sgn}(\theta_{b})\quad if \quad \theta_{b}\ne 0 \quad and \quad e_{b}\in [-1,1]\quad if \quad \theta_{b}=0\}$ and $S_{2}=\{e\in \mathbb{R}^{p(n)}: e_{b}=\alpha_{b}\quad if \quad \alpha_{b}\ne 0 \quad and \quad e_{b}\in [-1,1]\quad if \quad \alpha_{b}=0 \quad \mathrm{where} \quad \alpha = D'^{a}\mathrm{sgn}(D^{a}\theta)\}$. Observe that $|(D'^{a}\mathrm{sgn}(D^{a}\theta))_{b}| \leq \big\| D^{a}_{.b}\big\|_{1}$ where $D^{a}_{.b}$ denotes the $b^{th}$ column of the matrix $D^{a}$. This is same as the total number of local neighbors of $b$ except $a$. The lemma follows.
\end{proof}

\begin{proof}[{\rm \textbf{Proof of lemma~\ref{signconsist}}}]
This proof can be found in \ref{suppA}.
\end{proof}

\begin{proof}[{\rm \textbf{Proof of theorem~\ref{type1}}}]
The event $\hat{\mathrm{ne}}^{\lambda,\mu}_a \nsubseteq \mathrm{ne}_a$ is equivalent to the event that there exists some node $b\in \Gamma(n)\setminus\mathrm{cl}_a$ in the set of non-neighbors of node $a$ such that the estimated coefficient $\hat{\theta}^{a,\lambda,\mu}_b$ is not zero. Thus,
\begin{center}
$P\left(\hat{\mathrm{ne}}^{\lambda,\mu}_a \subseteq \mathrm{ne}_a\right) = 1 - P\left(\exists b \in \Gamma(n)\setminus\mathrm{cl}_a: \hat{\theta}^{a,\lambda,\mu}_b\neq 0\right).$
\end{center}
Let $\mathcal{E}$ be the event that
\begin{align*}
\max_{k\in\Gamma(n)\setminus\mathrm{cl}_a}\left|G_k\left(\hat{\theta}^{a,\mathrm{ne}_a,\mathcal{B},\lambda,\mu}\right)\right|<\lambda + B\mu.
\end{align*}
Conditional on $\mathcal{E}$, it follows from Meinshausen-B\"{u}hlmann's discussion that $\hat{\theta}^{a,\mathrm{ne}_a,\mathcal{B},\lambda,\mu}$ is also a solution to the fused LASSO problem with $\mathcal{A}=\Gamma(n)\setminus\{a\}$. As $\hat{\theta}^{a,\mathrm{ne}_a,\mathcal{B},\lambda,\mu}_b = 0$ for all $b\in\Gamma(n)\setminus\mathrm{cl}_a$, it follows that $\hat{\theta}^{a,\lambda,\mu}_b = 0$ for all $b\in\Gamma(n)\setminus\mathrm{cl}_a$. By~\ref{baselemma} ,
\begin{align*}
P\left(\exists b \in \Gamma(n)\setminus\mathrm{cl}_a: \hat{\theta}^{a,\lambda,\mu}_b\neq 0\right) \leq P\left(\max\limits_{k\in\Gamma(n)\setminus\mathrm{cl}_a}\left|G_k\left(\hat{\theta}^{a,\mathrm{ne}_a,\mathcal{B},\lambda,\mu}\right)\right|\geq\lambda + B\mu\right).
\end{align*}
It suffices to show that there exists a constant $c>0$ so that for all $b\in\Gamma(n)\setminus\mathrm{cl}_a$,
\begin{center}
$P\left(\left|G_b\left(\hat{\theta}^{a,\mathrm{ne}_a,\lambda,\mu}\right)\right|\geq \lambda + B\mu\right) = O(\exp(-cn^\varepsilon)).$
\end{center}
Writing $X_b = \sum_{m\in\mathrm{ne}_a}\theta^{b,\mathrm{ne}_a}_m X_m + V_b$ for any $b\in \Gamma(n)\setminus\mathrm{cl}_a$ where $V_b \sim N(0,\sigma^2_b)$ for some $\sigma^2_b\leq 1$ and $V_b$ is independent of $\{X_m;m\in\mathrm{cl}_a\}$. Hence,
\begin{align*}
G_b\left(\hat{\theta}^{a,\mathrm{ne}_a,\lambda,\mu}\right) = -2n^{-1}\sum_{m\in\mathrm{ne}_a} &\Big[\theta^{b,\mathrm{ne}_a}_m\langle X_a - X\hat{\theta}^{a,\mathrm{ne}_a,\lambda,\mu}, X_m\rangle \\
&- 2n^{-1}\langle X_a - X\hat{\theta}^{a,\mathrm{ne}_a,\lambda,\mu}, V_b\rangle\Big]
\end{align*}
By lemma~\ref{signconsist}, there exists a $c>0$ so that with probability $1-O(\exp(-cn^\varepsilon))$,
\begin{center}
$\mathrm{sgn}\left(\hat{\theta}^{a,\mathrm{ne}_a,\lambda,\mu}_k\right) = \mathrm{sgn}\left(\theta^{a,\mathrm{ne}_a}_k\right)$ $\forall k\in\mathrm{ne}_a.$
\end{center}
In this case, it holds by lemma~\ref{baselemma} that
\begin{align*}
\left|2n^{-1}\sum_{m\in\mathrm{ne}_a}\theta^{b,\mathrm{ne}_a}_m\langle X_a - X\hat{\theta}^{a,\mathrm{ne}_a,\lambda,\mu}, X_m\rangle \right| &\leq \left|\sum_{m\in\mathrm{ne}_a}\mathrm{sgn}\left(\theta^{a,\mathrm{ne}_a}_m\right)\theta^{b,\mathrm{ne}_a}_m \lambda\right| \\
&+ \left|\sum_{m\in\mathrm{ne}_a}[D'^a\mathrm{sgn}(D^a\theta)]_m\theta^{b,\mathrm{ne}_a}_m \mu\right| \\
&\leq \delta_1\lambda + \delta_2 B\mu
\end{align*}
The absolute value of the coefficient $G_b$ is hence bounded by
\begin{center}
$\left|G_b\left(\hat{\theta}^{a,\mathrm{ne}_a,\lambda,\mu}\right)\right| \leq \delta_1\lambda +\delta_2 B\mu + \left|2n^{-1}\langle X_a - X\hat{\theta}^{a,\mathrm{ne}_a,\lambda,\mu}, V_b \rangle\right|.$
\end{center}
with probability $1-O(\exp(-cn^{\varepsilon}))$. Conditional on $X_{\mathrm{cl}_a}$, the random variable $\langle X_a - X\hat{\theta}^{a,\mathrm{ne}_a,\lambda,\mu}_k, V_b\rangle$ is normally distributed with mean $0$ and variance $\sigma^2_b\|X_a - X\hat{\theta}^{a,\mathrm{ne}_a,\lambda,\mu}_k\|^2$. By definition of $\hat{\theta}^{a,\mathrm{ne}_a,\lambda,\mu}$,
\begin{center}
$\|X_a - X\hat{\theta}^{a,\mathrm{ne}_a,\lambda,\mu}_k\|^2 \leq \|X_a\|^2$
\end{center}
Since $\sigma^2_b \leq 1$, $2n^{-1}\langle X_a - X\hat{\theta}^{a,\mathrm{ne}_a,\lambda,\mu}_k, V_b\rangle$ is stochastically smaller than or equal to $\left|2n^{-1}\langle X_a,V_b\rangle\right|$. It remains to show that for some $c>0$ and some $0< \delta_1,\delta_2 <1$,
\begin{center}
$P\left(\left|2n^{-1}\langle X_a, V_b\rangle\right| \geq (1-\delta_1)\lambda + (1-\delta_2)B\mu\right) = O(\exp(-cn^{\varepsilon})).$
\end{center}
If $X_a$ and $V_b$ are independent, $E(X_a V_b) = 0$. Using Gaussianity and bounded variance of both $X_a$ and $V_b$, we obtain existence of some $g<\infty$ such that $E\left(\exp(|X_a V_b|)\right) < g$. Hence using Bernstein inequality and boundedness of $\lambda$ and $\mu$, it holds for some $c>0$ that for all $b\in\mathrm{ne}_a$ (see lemma~\ref{fastconv} for detailed proof),
\begin{equation}\label{unexplained}
P\left(\left|2n^{-1}\langle X_a, V_b\rangle\right| \geq (1-\delta_1)\lambda + (1-\delta_2)B\mu\right) = O(\exp(-cn^\varepsilon))
\end{equation}
which completes the proof.
\end{proof}

\begin{proof}[{\rm \textbf{Proof of proposition~\ref{type1problem}}}]
We follow the proof of theorem~\ref{type1} and claim that with the above assumptions, for all $a,b$ with $b\notin \mathrm{ne}_a$,
\begin{center}
$P\left(|G_b\left(\hat{\theta}^{a,\mathrm{ne}_a,\lambda,\mu}\right)|>\lambda+B\mu\right) \rightarrow 1$ for $n\rightarrow\infty$
\end{center}
Following similar arguments afterwards, it can be concluded that with some $\delta_1>1$ and $\delta_2>1$ as $n\rightarrow\infty$
\begin{center}
$P\left(|G_b\left(\hat{\theta}^{a,\mathrm{ne}_a,\lambda,\mu}\right)|\geq \delta_1\lambda + \delta_2 B\mu - |2n^{-1}\langle X_a - X\hat{\theta}^{a,\mathrm{ne}_a,\lambda,\mu},V_b \rangle|\right) \rightarrow 1.$ 
\end{center}
It holds for the third term that for any $g_1>0, g_2>0$,
\begin{center}
$P\left(|2n^{-1}\langle X_a - X\hat{\theta}^{a,\mathrm{ne}_a,\lambda,\mu},V_b\rangle| > g_1 \lambda + g_2 B\mu\right) \rightarrow 0$ as $n\rightarrow \infty.$
\end{center}
which combined with the previous result proves the proposition.
\end{proof}

\begin{proof}[{\rm \textbf{Proof of theorem~\ref{type2}}}]
Observe that
\begin{center}
$P\left(\mathrm{ne}_a \subseteq \hat{\mathrm{ne}}^\lambda_a\right) = 1 - P\left(\exists b\in \mathrm{ne}_a : \hat{\theta}^{a,\lambda,\mu}_b = 0\right).$
\end{center}
Let $\mathcal{E}$ be the event
\begin{center}
$\max_{k\in \Gamma(n)\setminus\mathrm{cl}_a}|G_k\left(\hat{\theta}^{a,\mathrm{ne}_a,\lambda,\mu}\right)| < \lambda + B\mu.$
\end{center}
On $\mathcal{E}$, following similar arguments as before, we can conclude that $\hat{\theta}^{a,\mathrm{ne}_a,\lambda,\mu} = \hat{\theta}^{a,\lambda,\mu}$. Therefore
\begin{center}
$P\left(\exists b\in \mathrm{ne}_a :\hat{\theta}^{a,\lambda,\mu}_b = 0\right) \leq P\left(b\in \mathrm{ne}_a : \hat{\theta}^{a,\mathrm{ne}_a,\lambda,\mu}=0\right) + P(\mathcal{E}^c).$
\end{center}
It follows from the proof of Theorem 4.6 that there exists some $c > 0$ so that
$P(\mathcal{E}^c) = O(\exp(-cn^\varepsilon))$. Using Bonferronis inequality, it hence remains to
show that there exists some $c > 0$ so that for all $b\in \mathrm{ne}_a$,
\begin{center}
$P\left(\hat{\theta}^{a,\mathrm{ne}_a,\lambda,\mu} = 0\right) = O(\exp(-cn^\varepsilon)),$
\end{center}
which follows from lemma~\ref{signconsist}.
\end{proof}

\begin{proof}[{\rm \textbf{Proof of theorem~\ref{regparam}}}]
$\hat{C}^{\lambda,\mu}_a \nsubseteq C_a$ implies the existence of an edge in the estimated neighborhood that connects two nodes in two different connectivity components of the true underlying graph. Hence,
\begin{center}
$P\left(\exists b\in\Gamma(n): b\in\hat{\mathrm{ne}}^{\lambda,\mu}_a\right) \leq p(n) \max_a P\left(\exists b\in \Gamma(n)\setminus C_a: b\in \hat{\mathrm{ne}}^{\lambda,\mu}_a\right)$
\end{center}
Going by the same arguments used in proving theorem 4.6, we have
\begin{center}
$P\left(\exists b\in \Gamma(n)\setminus C_a: b\in \hat{\mathrm{ne}}^{\lambda,\mu}_a\right) \leq P\left(\max_{b\in\Gamma(n)\setminus C_a} |G_b(\hat{\theta}^a,C_a,\lambda,\mu)| \geq \lambda + B\mu\right).$
\end{center}
Thus, it is sufficient to show that
\begin{center}
$p(n)^2 \max_{a\in\Gamma(n), b\in\Gamma(n)\setminus C_a} P\left(|G_b(\hat{\theta}^a,C_a,\lambda,\mu)| \geq \lambda + B\mu\right).$
\end{center}
Now observe that since $X_b$ and $\{X_k; k\in C_a\}$ are in different connectivity component, they are, in fact, independent. Therefore, conditional on $X_{C_a}$, $G_b(\hat{\theta}^a,C_a,\lambda,\mu) \sim N\left(0, \frac{4\|X_a - X\hat{\theta}^{a,C_a,\lambda,\mu}\|^2}{n^2}\right)$, making it stochastically smaller than $Z \sim N\left(0,\frac{4\|X_a\|^2}{n^2}\right)$. Hence it holds for all $a\in \Gamma(n)$ and $b\in\Gamma(n)\setminus C_a$ that
\begin{center}
$P\left(|G_b(\hat{\theta}^{a,C_a,\lambda,\mu})|>\lambda+B\mu\right) \leq 2\tilde{\Phi}\left(\frac{\sqrt{n}(\lambda+B\mu)}{2\hat{\sigma_a}}\right),$
\end{center}
where $\tilde{\Phi} = 1 - \Phi$. Using the $\lambda$ and $\mu$ proposed, the RHS becomes $\frac{\alpha}{p(n)^2}$.
\end{proof}

\begin{proof}[{\rm \textbf{Proof of lemma~\ref{fastconv}}}]
This is a direct application of Bernstein inequality. Since $\lambda\rightarrow 0$ and $\mu\rightarrow 0$ as $n\rightarrow\infty$, for large enough $n$, we have, by a version of Bernstein's inequality, that
\begin{align*}
 &P\left(\left| 2n^{-1}\langle X_a, V_b\rangle \right| \geq (1-\delta_1)\lambda + (1-\delta_2)B\mu\right)\\
& \hspace*{3cm}\leq \exp\left[-\frac{\left(\frac{d_1}{2}(1-\delta_1)n^{\frac{1+\epsilon}{2}} + \frac{d_2}{2}(1-\delta_2)n^{\frac{1+\epsilon}{2} + \beta_0} \right)^2}{nE(X^2_{a,i} V^2_{b,i})}\right]\\
   & \hspace*{3cm}= \exp\left[-\frac{1}{\sigma^2_*} \left(\frac{d_1}{2}(1 - \delta_1)+ \frac{d_2}{2}(1 - \delta_2) n^{\beta_0}\right)^2 n^\epsilon\right].\\
    \intertext{This proves part (a). Write the last expression as}
 &e^{-\frac{1}{\sigma^2_*}\left(\frac{d^2_1}{4}(1-\delta_1)^2 n^{\epsilon}\right)} \cdot e^{-\frac{1}{\sigma^2_*}\left(\frac{d_1 d_2}{2}(1 - \delta_1)(1 - \delta_2)n^{\beta_0} + \frac{d^2_2}{4}(1 - \delta_2)^2 n^{2\beta_0}\right)n^{\epsilon}}
    \intertext{proves part (b), since the first factor is the upper bound for lasso.}
    \end{align*}
    
    \vspace*{-1cm}
\end{proof}

\begin{proof}[{\rm \textbf{Proof of lemma~\ref{basicineq}}}]
Since $\hat{\theta}^{\lambda,\mu}_a$ is the fused lasso minimizer, we get
\begin{align*}
&\frac{1}{n}\big\| X_a - X^a \hat{\theta}^{\lambda,\mu}_a\big\|^2 + \lambda\big\|\hat{\theta}^{\lambda,\mu}_a\big\|_1 + \mu\big\| D^a\hat{\theta}^{\lambda,\mu}_a\big\|_1 \\
& \hspace*{4cm}\leq \frac{1}{n}\big\| X_a - X^a \theta^0_a\big\|^2 + \lambda\big\|\theta^0_a\big\|_1 + \mu\big\| D^a\theta^0_a\big\|_1.
\intertext{Plugging in $X_a = X^a \theta^0_a + \epsilon_a$, we further have}
& \frac{1}{n}\big\| X^a( \hat{\theta}^{\lambda,\mu}_a - \theta^0_a\big\|^2 - \frac{2}{n} \epsilon'_a X^a(\hat{\theta}^{\lambda,\mu}_a - \theta^0_a) + \frac{1}{n}\big\|\epsilon_a\big\|^2 + \lambda\big\|\hat{\theta}^{\lambda,\mu}_a\big\|_1 + \mu\big\| D^a\hat{\theta}^{\lambda,\mu}_a\big\|_1\\
&\hspace{5cm}\leq \frac{1}{n}\big\|\epsilon_a\big\|^2 + \lambda\big\|\theta^0_a\big\|_1 + \mu\big\| D^a\theta^0_a\big\|_1.\\
&\text{Rewrite the above inequality to get the desired lemma.}
\end{align*}

\vspace*{-0.5cm}
\end{proof}

\begin{proof}[{\rm \textbf{Proof of lemma~\ref{highprob}}}]
The proof can be found in \ref{suppA}.
\end{proof}

\begin{proof}[{\rm \textbf{Proof of lemma~\ref{l2ineq}}}]
The proof exploits the fact that $\hat{\theta}^{\lambda,\mu}_a$ is the minimizer of the penalized least square by equalling the sub-differential of the objective function to 0. We start by replacing the assumed linear truth, i.e., $X_a = X^a\theta^0_a + \epsilon_a$. Therefore, we get by using the notation ${\mathbb D}^a = D^{a'}\mathrm{sgn}(D^a \hat{\theta}^{\lambda,\mu}_a),$ and $(X^a)^2 = X^{a^\prime}X^a$ that
\begin{align*}
&\frac{\partial}{\partial\theta_a} \left[\frac{1}{n}\big\| X^a (\theta^0_a - \theta_a) \big\|^2 + \frac{1}{n}\big\|\epsilon_a\big\|^2 + \frac{2}{n}\epsilon'_a X^a\left(\theta^0_a - \theta_a\right) \right.+\\
&\hspace*{6cm} \left.\lambda \mathrm{sgn}(\theta_a) + \mu {\mathbb D}^a\right]_{\theta_a = \hat{\theta}^{\lambda,\mu}_a} = 0\\
&\Rightarrow\quad \frac{2}{n}\left((X^a)^2\,( \hat{\theta}^{\lambda,\mu}_a - \theta_0) \right) - \frac{2}{n}X^{a'}\epsilon_a + \lambda\mathrm{sgn}\left(\hat{\theta}^{\lambda,\mu}_a\right) + \mu {\mathbb D}^a  = 0\\
&\Rightarrow\quad (X^a)^2 \hat{\theta}^{\lambda,\mu}_a =  (X^a)^2\theta_0 + \frac{n}{2}\left(\frac{2}{n}X^{a'}\epsilon_a\right) - \frac{n\lambda}{2}\mathrm{sgn}\left(\hat{\theta}^{\lambda,\mu}_a\right) - \frac{n\mu}{2} {\mathbb D}^a \\
&\Rightarrow\quad \hat{\theta}^{\lambda,\mu}_a = \left((X^a)^2\right)^+ (X^a)^2 \theta^0_a + \frac{n}{2}\left((X^a)^2\right)^+ \left(\frac{2}{n}X^{a'}\epsilon_a - \lambda\mathrm{sgn}\left(\hat{\theta}^{\lambda,\mu}_a\right) - \mu {\mathbb D}^a\right).
\end{align*}
With the given choices for $\lambda$ and $\mu$, define $\Lambda_a = \Big\{\mathrm{max}_{\stackrel{1\leq j\leq p}{j\ne a}} \frac{2}{n} \left|\epsilon'_a X^a_j\right| \leq \lambda + B\mu\Big\}$, then we have, with probability $1 - \exp(-t^2)$ that
\begin{align*}
\left|\frac{2}{n}X^{a'}\epsilon_a - \lambda\mathrm{sgn}\left(\hat{\theta}^{\lambda,\mu}_a\right) - \mu {\mathbb D}^a\right| \leq  (\lambda+B\mu)\mathbf{1}_p + \lambda\mathbf{1}_p + B\mu \mathbf{1}_p = 2(\lambda + B\mu)\mathbf{1}_p,
\end{align*}
where the inequality is meant to be interpreted componentwise and $\mathbf{1}_p$ is a vector of size $p$ consisting of 1's. Hence, we get
\begin{align*}
&\big\|\hat{\theta}^{\lambda,\mu}_a\big\|_1  = \left|\left|(X^a)^{2^+} (X^a)^2 \theta^0_a + \frac{n}{2}(X^a)^{2^+} \left(\frac{2}{n}X^{a'}\epsilon_a - \lambda\mathrm{sgn}\left(\hat{\theta}^{\lambda,\mu}_a\right) - \mu {\mathbb D}^a \right)\right|\right|_1\\
&\leq \left|\left|(X^a)^{2^+} (X^a)^2 \theta^0_a\right|\right|_1 + \frac{n}{2}\left|\left|\left(X^{a'}X^a\right)^+ \left(\frac{2}{n}X^{a'}\epsilon_a - \lambda\mathrm{sgn}\left(\hat{\theta}^{\lambda,\mu}_a\right) - \mu {\mathbb D}^a \right)\right|\right|_1\\
&\leq \sqrt{p}\left|\left|(X^a)^{2^+} (X^a)^2 \theta^0_a\right|\right|_2 + \frac{n\sqrt{p}}{2}\left|\left|(X^a)^{2^+} \left(\frac{2}{n}X^{a'}\epsilon_a - \lambda\mathrm{sgn}\left(\hat{\theta}^{\lambda,\mu}_a\right) - \mu {\mathbb D}^a \right)\right|\right|_2
\end{align*}
Using the facts that $\big\| Ax\big\|_2 \leq s_{\mathrm{max}} \big\| x\big\|_2$, where $s_{\mathrm{max}}$ is the maximum singular value of $A$ and that $A^+ A$ is idempotent, and hence its singular values are either 0 or 1, we can continue the above sequence of inequalities and obtain 
\begin{align*}
&\hspace*{1.5cm}\leq \sqrt{p}\big\|\theta^0_a\big\|_2 + \frac{n\sqrt{p}}{2\delta_{\mathrm{min}}}\left|\left|\frac{2}{n}X^{a'}\epsilon_a - \lambda\mathrm{sgn}\left(\hat{\theta}^{\lambda,\mu}_a\right) - \mu {\mathbb D}^a\right|\right|_2\\
&\hspace*{1.5cm}\leq \sqrt{p}\big\|\theta^0_a\big\|_2 + \frac{np}{2\delta_{\mathrm{min}}}\left|\left|\frac{2}{n}X^{a'}\epsilon_a - \lambda\mathrm{sgn}\left(\hat{\theta}^{\lambda,\mu}_a\right) - \mu {\mathbb D}^a\right|\right|_\infty\\
&\hspace*{1.5cm}\leq \sqrt{p}\big\|\theta^0_a\big\|_2 + \frac{np(\lambda + B\mu)}{\delta_{\mathrm{min}}}
\end{align*}
Plugging in the result obtained from previous corollary, we get
\begin{align*}
&\textstyle{P\left(\frac{1}{n}\big\| X^a (\hat{\theta}^{\lambda,\mu}_a - \theta^0_a)\big\|^2 \leq 2(\lambda + B\mu)\big\| \theta^0_a \big\|_1 + \mu B\left(\sqrt{p}\big\|\theta^0_a\big\|_2 + \frac{np(\lambda + B\mu)}{\delta_{\mathrm{min}}}\right) \right) \geq 1 - e^{-t^2}}
\end{align*}
which implies that
\begin{align*}
{\textstyle P\left(\frac{1}{n}\big\| X^a (\hat{\theta}^{\lambda,\mu}_a - \theta^0_a)\big\|^2 \leq 2\left(\lambda + B\mu\left(1 + \frac{\sqrt{p}}{2}\right)\right)\big\| \theta^0_a \big\|_1 + \frac{npB\mu(\lambda + B\mu)}{\delta_{\mathrm{min}}}\right) \geq 1 - e^{-t^2}.}
\end{align*}
\end{proof}

\begin{proof}[{\rm \textbf{Proof of lemma~\ref{oracle1}}}]
To start with, we derive a series of inequalities and apply them on the basic inequality.
\begin{align*}
&\big\|\hat{\theta}^{\lambda,\mu}_a\big\|_1= \big\|\hat{\theta}^{\lambda,\mu}_{a,S_0}\big\|_1 + \big\|\hat{\theta}^{\lambda,\mu}_{a,S^c_0}\big\|_1 \geq \big\|\theta^0_{a,S_0}\big\|_1 - \big\|\hat{\theta}^{\lambda,\mu}_{a,S_0} - \theta^0_{a,S_0}\big\|_1 + \big\|\hat{\theta}^{\lambda,\mu}_{a,S^c_0}\big\|_1\\
&\big\|\hat{\theta}^{\lambda,\mu}_a - \theta^0_a\big\|_1 \quad = \quad \big\|\hat{\theta}^{\lambda,\mu}_{a,S_0} - \theta^0_{a,S_0}\big\|_1 + \big\|\hat{\theta}^{\lambda,\mu}_{a,S_0}\big\|_1
\end{align*}
We write $\hat{\theta}^{\lambda,\mu}_{a,S_0} = \left(\hat{\theta}^{\lambda,\mu}_{a,S_0,1},0\right)'$ and $\hat{\theta}^{\lambda,\mu}_{a,S^c_0} = \left(0, \hat{\theta}^{\lambda,\mu}_{a,S^c_0,1}\right)'$ so that $\hat{\theta}^{\lambda,\mu}_a = \left(\hat{\theta}^{\lambda,\mu}_{a,S_0,1},\hat{\theta}^{\lambda,\mu}_{a,S^c_0,1}\right)'$. Similarly we write $\theta^0_a = \theta^0_{a,S_0} = \left(\theta^0_{a,S_0,1},0\right)'$. Hence, we get
\begin{align*}
&\big\| D^a\hat{\theta}^{\lambda,\mu}_a\big\|_1= 
\left|\left|\left[
\begin{matrix}
D^a_{S_0,S_0} & 0\\
D^a_{S_0,0} & D^a_{0,S^c_0}\\
0 & D^a_{S^c_0,S^c_0}
\end{matrix}
\right]
\left(
\begin{matrix}
\hat{\theta}^{\lambda,\mu}_{a,S_0,1}\\
\hat{\theta}^{\lambda,\mu}_{a,S^c_0,1}
\end{matrix}
\right)\right|\right|_1 = 
\left|\left|
\left(
\begin{matrix}
D^a_{S_0,S_0} \hat{\theta}^{\lambda,\mu}_{a,S_0,1}\\
D^a_{S_0,0} \hat{\theta}^{\lambda,\mu}_{a,S_0,1} + D^a_{0,S^c_0} \hat{\theta}^{\lambda,\mu}_{a,S^c_0,1} \\
D^a_{S^c_0,S^c_0} \hat{\theta}^{\lambda,\mu}_{a,S^c_0,1}
\end{matrix}
\right)
\right|\right|_1
\end{align*}
\begin{align*}
&\geq\quad
 \big\| D^a_{S_0,S_0} \hat{\theta}^{\lambda,\mu}_{a,S_0,1}\big\|_1 + \big\| D^a_{S^c_0,S^c_0} \hat{\theta}^{\lambda,\mu}_{a,S^c_0,1}\big\|_1 + \big\| D^a_{S_0,0} \hat{\theta}^{\lambda,\mu}_{a,S_0,1}\big\|_1 - \big\| D^a_{0,S^c_0} \hat{\theta}^{\lambda,\mu}_{a,S^c_0,1}\big\|_1\\
&\geq\quad \big\| D^a_{S_0,S_0} \theta^0_{a,S_0,1}\big\|_1 - \big\| D^a_{S_0,S_0}\left( \hat{\theta}^{\lambda,\mu}_{a,S_0,1} - \theta^0_{a,S_0,1}\right)\big\|_1 + \big\| D^a_{S^c_0,S^c_0} \hat{\theta}^{\lambda,\mu}_{a,S^c_0,1}\big\|_1 \\
&\hspace*{2cm} + \big\| D^a_{S_0,0} \theta^0_{a,S_0,1}\big\|_1 - \big\| D^a_{S_0,0} \left( \hat{\theta}^{\lambda,\mu}_{a,S_0,1} - \theta^0_{a,S_0,1}\right)\big\|_1 - \big\| D^a_{0,S^c_0} \hat{\theta}^{\lambda,\mu}_{a,S^c_0,1} \big\|_1\\
&\geq \quad \big\| D^a_{S_0,S_0} \theta^0_{a,S_0,1}\big\|_1 + \big\| D^a_{S_0,0} \theta^0_{a,S_0,1}\big\|_1 - 2B\big\| \hat{\theta}^{\lambda,\mu}_{a,S_0} - \theta^0_{a,S_0}\big\|_1 \\
&\hspace*{2cm}+ \big\| D^a_{S^c_0,S^c_0} \hat{\theta}^{\lambda,\mu}_{a,S^c_0,1}\big\|_1 - B \big\|\hat{\theta}^{\lambda,\mu}_{a,S^c_0} \big\|_1
\end{align*}
Similarly, we get
\begin{align*}
\big\| D^a\theta^0_a\big\|_1 \quad &= \quad
\left|\left|\left[
\begin{matrix}
D^a_{S_0,S_0} & 0\\
D^a_{S_0,0} & D^a_{0,S^c_0}\\
0 & D^a_{S^c_0,S^c_0}
\end{matrix}
\right]
\left(
\begin{matrix}
\theta^0_{a,S_0,1}\\
0
\end{matrix}
\right)\right|\right|_1 \quad = \quad
\left|\left|
\left(
\begin{matrix}
D^a_{S_0,S_0} \theta^0_{a,S_0,1}\\
D^a_{S_0,0} \theta^0_{a,S_0,1}\\
0
\end{matrix}
\right)\right|\right|_1\\
&= \quad \big\| D^a_{S_0,S_0} \theta^0_{a,S_0,1}\big\|_1 + \big\| D^a_{S_0,0} \theta^0_{a,S_0,1} \big\|_1
\end{align*}
Plugging into the basic inequality, we get on $\Lambda_a$, with $\lambda\geq 2\lambda_0$ and $\mu\geq 2\mu_0$,
\begin{align*}
&\frac{1}{n}\big\| X^a \left(\hat{\theta}^{\lambda,\mu}_a - \theta^0_a\right)\big\|^2 +  \lambda\big\|\theta^0_{a,S_0}\big\|_1 - \lambda\big\|\hat{\theta}^{\lambda,\mu}_{a,S_0} - \theta^0_{a,S_0}\big\|_1 + \lambda\big\|\hat{\theta}^{\lambda,\mu}_{a,S^c_0}\big\|_1\\
& + \mu\big\| D^a_{S_0,S_0} \theta^0_{a,S_0,1}\big\|_1 + \mu\big\| D^a_{S_0,0} \theta^0_{a,S_0,1}\big\|_1 - 2B\mu\big\| \hat{\theta}^{\lambda,\mu}_{a,S_0} - \theta^0_{a,S_0}\big\|_1 \\
& + \mu\big\| D^a_{S^c_0,S^c_0} \hat{\theta}^{\lambda,\mu}_{a,S^c_0,1}\big\|_1 - B\mu\big\| \hat{\theta}^{\lambda,\mu}_{a,S^c_0} \big\|_1\\
&\vspace{.5cm}\\
\leq\hspace{1cm} & \frac{\lambda+B\mu}{2}\big\|\hat{\theta}^{\lambda,\mu}_{a,S_0} - \theta^0_{a,S_0}\big\|_1 + \frac{\lambda+B\mu}{2}\big\|\hat{\theta}^{\lambda,\mu}_{a,S^c_0}\big\|_1 + \lambda\big\| \theta^0_{a,S_0}\big\|_1 \\
& + \mu\big\| D^a_{S_0,S_0} \theta^0_{a,S_0,1}\big\|_1 + \mu\big\| D^a_{S_0,0} \theta^0_{a,S_0,1} \big\|_1.
\end{align*}
From this, we get
\begin{align*}
&\frac{2}{n}\big\| X^a \left(\hat{\theta}^{\lambda,\mu}_a - \theta^0_a\right)\big\|^2 + (\lambda - 3B\mu) \big\|\hat{\theta}^{\lambda,\mu}_{a,S^c_0}\big\|_1 \quad\leq\quad (3\lambda+5B\mu) \big\|\hat{\theta}^{\lambda,\mu}_{a,S_0} - \theta^0_{a,S_0}\big\|_1
\end{align*}
\end{proof}

\begin{proof}[{\rm \textbf{Proof of lemma~\ref{oracle2}}}]
The proof is basically a continuation of what we have already shown. We have
\begin{align*}
& \qquad\frac{2}{n}\big\| X^a \left(\hat{\theta}^{\lambda,\mu}_a - \theta^0_a\right)\big\|^2 + (\lambda - 3B\mu) \big\|\hat{\theta}^{\lambda,\mu}_a - \theta^0_a\big\|_1 \\
&= \quad \frac{2}{n}\big\| X^a \left(\hat{\theta}^{\lambda,\mu}_a - \theta^0_a\right)\big\|^2 + (\lambda - 3B\mu) \big\|\hat{\theta}^{\lambda,\mu}_{a,S_0} - \theta^0_{a,S_0}\big\|_1 + (\lambda - 3B\mu)\big\|\hat{\theta}^{\lambda,\mu}_{a,S^c_0}\big\|_1\\
&\leq \quad (3\lambda+5B\mu)\big\|\hat{\theta}^{\lambda,\mu}_{a,S_0} - \theta^0_{a,S_0}\big\|_1 + (\lambda - 3B\mu)\big\|\hat{\theta}^{\lambda,\mu}_{a,S_0} - \theta^0_{a,S_0}\big\|_1 \\
&= \quad 2(2\lambda+B\mu)\big\|\hat{\theta}^{\lambda,\mu}_{a,S_0} - \theta^0_{a,S_0}\big\|_1 \quad \leq \quad \frac{2\sqrt{s_0}(2\lambda+B\mu)}{\sqrt{n}\phi_{0,a}}\big\|\hat{\theta}^{\lambda,\mu}_{a,S_0} - \theta^0_{a,S_0}\big\|\\
&\leq \quad \frac{s_0(2\lambda+B\mu)^2}{\phi^2_{0,a}} + \frac{1}{n}\big\| X^a \left(\hat{\theta}^{\lambda,\mu}_a - \theta^0_a\right)\big\|^2
\end{align*}

\end{proof}

\begin{supplement}[id=suppA]
  \sname{Supplement A}
  \stitle{Proofs of some results}
  \sdescription{This supplement contains the proofs of some of the results.}
\end{supplement}

\section*{Acknowledgements}
The authors are thankful to Dr. Debashis Paul and Dr. Ethan Anderes who helped shape some ideas presented in this paper with insightful comments and discussions.

\vfill

\newpage


\section*{Supplementary material}

# Supplementary material A: Proofs

**Proof of proposition 6.1.** The subdifferential of $E(X_a - \sum_{k \in \Gamma(n)})^2 + \eta_1 \|\theta\|_1 + \eta_2 \|D^a \theta^a\|_1$, w.r.t. $\theta_k^a$: $k \in \Gamma(n)$ is given by

$$-2E((X_a - \sum_{m \in \Gamma(n)} \theta_m^a X_m) X_k) + \eta_1 e_k^{(1)} + \eta_2 e_k^{(2)}$$

where $e_k^{(1)} = \text{sgn}(\theta_k^a)$ if $\theta_k^a \neq 0$ and $e_k^{(1)} \in [-1,1]$ if $\theta_k^a = 0$ and $e_k^{(2)} = (D'^a \text{sgn}(D^a \theta^a))_k$ if $(D'^a \text{sgn}(D^a \theta^a))_k \neq 0$. Otherwise, $e_k^{(2)} \in [-1,1]$. Using $\text{ne}_a(\eta_1, 0) = \text{ne}_a(0,0)$, it follows from lemma 6.2 that for all $k \in \text{ne}_a$,

$$2E((X_a - \sum_{m \in \Gamma(n)} \theta_m^a(\eta_1,0) X_m) X_k) = \eta_1 \text{sgn}(\theta_k^a)$$

and for $k \notin \text{ne}_a$,

$$|2E((X_a - \sum_{m \in \Gamma(n)} \theta_m^a(\eta_1,0) X_m) X_k)| \leq \eta_1$$

A variable $X_b$ with $b \notin \text{ne}_a$ can be written as $X_b = \sum_{k \in \text{ne}_a} \theta_k^{b,\text{ne}_a} X_k + W_b$, where $W_b$ is independent of $\{X_k : k \in \text{cl}_a\}$. Using this yields

$$|2 \sum_{k \in \text{ne}_a} \theta_k^{b,\text{ne}_a} E((X_a - \sum_{m \in \Gamma(n)} \theta_m^a(\eta_1,0) X_m) X_k)| \leq \eta_1$$

Thus, it follows that $|\sum_{k \in \text{ne}_a} \theta_k^{b,\text{ne}_a} \text{sgn}(\theta_k^{a,\text{ne}_a})| \leq 1$. Using $\text{ne}_a(0, \eta_2) = \text{ne}_a(0,0)$, it follows from lemma 6.2 that for all $k \in \mathcal{L}_a$,

$$2E((X_a - \sum_{m \in \Gamma(n)} \theta_m^a(0,\eta_2) X_m) X_k) = \eta_2 (D'^a \text{sgn}(D^a \theta^a))_k$$

and for $k \notin \mathcal{L}_a$,

$$|2E((X_a - \sum_{m \in \Gamma(n)} \theta_m^a(0,\eta_2) X_m) X_k)| \leq \eta_2 \|D_{.b}^a\|_1$$

A variable $X_b$ with $b \notin \mathcal{L}_a$ can be written as $X_b = \sum_{k \in \mathcal{L}_a} \theta_k^{b,\mathcal{L}_a} X_k + Z_b$, where $Z_b$ is independent of $\{X_k : k \in \mathcal{L}_a\}$. Using this yields

$$|2 \sum_{k \in \mathcal{L}_a} \theta_k^{b,\mathcal{L}_a} E((X_a - \sum_{m \in \Gamma(n)} \theta_m^a(0,\eta_2) X_m) X_k)| \leq \eta_2 \|D_{.b}^a\|_1$$

Thus, it follows that

$$|\sum_{k \in \mathcal{L}_a} \theta_k^{b,\mathcal{L}_a} (D'^a \text{sgn}(D^a \theta^a))_k| \leq \|D_{.b}^a\|_1$$

$\square$

**Proof of lemma 6.3.** Using Bonferronis inequality, and $|ne_a| = o(n)$ for $n \to \infty$, it suffices to show that there exists some $c > 0$ so that for for every $a, b \in \Gamma(n)$ with $b \in ne_a$, $\mathbb{P}(\text{sgn}(\theta_b^{a,ne_a,\mathcal{B},\lambda,\mu}) = \text{sgn}(\theta_b^a)) = 1 - O(\exp(-cn^\epsilon))$.

Consider the definition of

$$\hat{\theta}^{a,ne_a,\mathcal{B},\lambda,\mu} =$$
$$\operatorname{argmin}_{\theta:\theta_k=0\forall k\notin ne_a,\theta_l-\theta_m=0\forall(l,m)\in\mathcal{B}} \left(\tfrac{1}{n}\parallel X_a - X\theta \parallel_2^2 +\lambda\parallel\theta\parallel_1 +\mu\parallel D^a\theta\parallel_1\right)$$

Assume now that component $b$ of this estimate is fixed at a constant value $\beta$. Denote this new estimate by $\tilde{\theta}^{a,b,\mathcal{B},\lambda,\mu}(\beta)$,

$$\tilde{\theta}^{a,b,\mathcal{B},\lambda,\mu}(\beta) = \operatorname{argmin}_{\theta\in\Theta_{a,b}(\beta)}(n^{-1}\parallel X_a - X^a\theta\parallel_2^2 +\lambda\parallel\theta\parallel_1 +\mu\parallel D^a\theta\parallel_1)$$

where

$$\Theta_{a,b}(\beta) = \{\theta \in \mathbb{R}^{p(n)} : \theta_b = \beta, \theta_k = 0\forall k\notin ne_a, \theta_l - \theta_m = 0\forall(l,m)\in\mathcal{B}\}$$

There will always exist a value $\beta = \hat{\theta}_b^{a,ne_a,\mathcal{B},\lambda,\mu}$ such that $\tilde{\theta}^{a,b,\mathcal{B},\lambda,\mu}(\beta)$ is identical to $\hat{\theta}^{a,ne_a,\mathcal{B},\lambda,\mu}$. Thus, if $\operatorname{sgn}(\hat{\theta}_b^{a,ne_a,\mathcal{B},\lambda,\mu}) \neq \operatorname{sgn}(\theta_b^a)$, there would exist some $\beta$ with $\operatorname{sgn}(\beta)\operatorname{sgn}(\theta_b^a) \leq 0$ so that $\tilde{\theta}^{a,b,\mathcal{B},\lambda,\mu}(\beta)$ would be a solution. Using $\operatorname{sgn}(\theta_b^a) \neq 0$ for all $b \in ne_a$, it is sufficient to show that for every $\beta$ with $\operatorname{sgn}(\beta)\operatorname{sgn}(\theta_b^a) < 0$, $\tilde{\theta}^{a,b,\mathcal{B},\lambda,\mu}(\beta)$ can not be a solution with high probability.

We concentrate on the case where $\theta_b^a > 0$. The case $\theta_b^a < 0$ will follow analogously. If $\theta_b^a > 0$, it follows by lemma 1 that $\tilde{\theta}^{a,b,\mathcal{B},\lambda,\mu}(\beta)$ with $\tilde{\theta}_b^{a,b,\mathcal{B},\lambda,\mu}(\beta) = \beta \leq 0$ can only be a solution if $G_b(\tilde{\theta}^{a,b,\mathcal{B},\lambda,\mu}(\beta)) \geq -\lambda - B\mu$ where $B = \max_{a,b}\parallel D_{\cdot,b}^a\parallel_1$. Hence it suffices to show that for some $c > 0$ and all $b \in ne_a$ with $\theta_b^a > 0$, for $n \to \infty$

$$\mathbb{P}(\sup_{\beta\leq 0}\{G_b(\tilde{\theta}^{a,b,\mathcal{B},\lambda,\mu}(\beta))\} < -\lambda - B\mu) = 1 - O(\exp(-cn^\epsilon)) \qquad (0.1)$$

Let in the following $R_a^{\lambda,\mu}(\beta)$ be the $n$-dimensional vector of residuals.

$$R_a^{\lambda,\mu}(\beta) = X_a - X\tilde{\theta}^{a,b,\mathcal{B},\lambda,\mu}(\beta)$$

We can write $X_b$ as

$$X_b = \sum_{k\in ne_a\setminus\{b\}} \theta_k^{b,ne_a\setminus\{b\}} X_k + W_b$$

where $W_b$ is independent of $\{X_k : k \in ne_a\setminus\{b\}\}$. It follows that

$$G_b(\tilde{\theta}^{a,b,\mathcal{B},\lambda,\mu}(\beta)) =$$
$$-2n^{-1}\langle R_a^{\lambda,\mu}(\beta), W_b\rangle - \sum_{k\in ne_a\setminus\{b\}} \theta_k^{b,ne_a\setminus\{b\}}(2n^{-1}\langle R_a^{\lambda,\mu}(\beta), X_k\rangle)$$

By lemma 6.2, $\forall k \in ne_a\setminus\{b\}$, $|G_k(\tilde{\theta}^{a,b,\mathcal{B},\lambda,\mu}(\beta))| = |2n^{-1}\langle R_a^{\lambda,\mu}(\beta), X_k\rangle| \leq \lambda + B\mu$. This together with the equation above yields

$$G_b(\tilde{\theta}^{a,b,\mathcal{B},\lambda,\mu}(\beta)) \leq -2n^{-1}\langle R_a^{\lambda,\mu}(\beta), W_b\rangle + (\lambda + B\mu)\parallel \theta^{b,ne_a\setminus\{b\}}\parallel_1$$

Using Assumption 5, there exists some $\vartheta < \infty$, so that $\parallel \theta^{b,ne_a\setminus\{b\}}\parallel_1 < \vartheta$. It is therefore sufficient to show that there exists for every $g > 0$ some $c > 0$ so that it holds for all $b \in ne_a$ with $\theta_b^a > 0$, for $n \to \infty$,

$$\mathbb{P}(\inf_{\beta\leq 0}\{2n^{-1}\langle R_a^{\lambda,\mu}(\beta), W_b\rangle\} > g(\lambda + B\mu)) = 1 - O(\exp(-cn^\epsilon))$$

Let $\mathbf{W}^{||} \subseteq \mathbb{R}^n$ be the space spanned by the vectors $\{X_k, k \in ne_a \setminus \{b\}\}$ and let $\mathbf{W}^\perp$ be the orthogonal complement of $\mathbf{W}^{||}$ in $\mathbb{R}^n$. Split the n-dimensional vector $W_b$ into the two vectors $W_b = W_b^\perp + W_b^{||}$. where $W_b^{||} \in \mathbf{W}^{||}$ and $W_b^\perp \in \mathbf{W}^\perp$. The inner product can be written as

$$2n^{-1}\langle R_a^{\lambda,\mu}(\beta), W_b\rangle = 2n^{-1}\langle R_a^{\lambda,\mu}(\beta), W_b^{||}\rangle + 2n^{-1}\langle R_a^{\lambda,\mu}(\beta), W_b^\perp\rangle$$

By Lemma 0.1 (see below), there exists for every $g > 0$ some $c > 0$ so that, for $n \to \infty$

$$\mathbb{P}(\inf_{\beta \leq 0}\{2n^{-1}\langle R_a^{\lambda,\mu}(\beta), W_b^{||}\rangle/(1 + Kn^{\beta_0}|\beta|)\} > -g(\lambda + B\mu)) = 1 - O(\exp(-cn^\epsilon))$$

To show the result, it is sufficient to prove that there exists for every $g > 0$ some $c > 0$ so that, for $n \to \infty$,

$$\mathbb{P}(\inf_{\beta \leq 0}\{2n^{-1}\langle R_a^{\lambda,\mu}(\beta), W_b^\perp\rangle - g(1 + Kn^{\beta_0}|\beta|)(\lambda + B\mu)\} > g(\lambda + B\mu)) = 1 - O(\exp(-cn^\epsilon))$$

It holds for some random variable $V_a$, independent of $X_{ne_a}$, that

$$X_a = \sum_{k \in ne_a} \theta_k^a X_k + V_a$$

Note that $V_a$ and $W_b$ are independent normally distributed random variables with variances $\sigma_a^2$ and $\sigma_b^2$ respectively. By assumption 2, $0 < v^2 \leq \sigma_b^2, \sigma_a^2 \leq 1$. Note furthermore that $W_b$ and $X_{ne_a \setminus \{b\}}$ are independent. Using $\theta^a = \theta^{a,ne_a}$ and $X_b = \sum_{k \in ne_a \setminus \{b\}} \theta_k^{b,ne_a \setminus \{b\}} X_k + W_b$

$$X_a = \sum_{k \in ne_a \setminus \{b\}} (\theta_k^a + \theta_b^a \theta_k^{b,ne_a \setminus \{b\}})X_k + \theta_b^a W_b + V_a$$

Using this, the definition of residuals and the orthogonality property of $W_b^\perp$,

$$2n^{-1}\langle R_a^{\lambda,\mu}(\beta), W_b^\perp\rangle = 2n^{-1}(\theta_b^a - \beta)\langle W_b^\perp, W_b^\perp\rangle + 2n^{-1}\langle V_a, W_b^\perp\rangle \geq 2n^{-1}(\theta_b^a - \beta)\langle W_b^\perp, W_b^\perp\rangle - 2n^{-1}|\langle V_a, W_b^\perp\rangle|$$

The second term, $2n^{-1}|\langle V_a, W_b^\perp\rangle|$, is stochastically smaller than $2n^{-1}|\langle V_a, W_b\rangle|$. Due to independence of $V_a$ and $W_b$, $E(V_a W_b) = 0$. Using Bernstein inequality, and $\lambda + B\mu \sim dn^{-\frac{1-\epsilon}{2}}$ with $\epsilon > 0$, there exists for every $g > 0$, some $c > 0$ so that

$$P(2n^{-1}|\langle V_a, W_b^\perp\rangle| \geq g(\lambda + B\mu)) \leq P(2n^{-1}|\langle V_a, W_b\rangle| \geq g(\lambda + B\mu)) = O(\exp(-cn^\epsilon))$$

Thus, it is sufficient to show that for every $g > 0$, there exists a $c > 0$ such that for $n \to \infty$,

$$P(\inf_{\beta \leq 0}\{2n^{-1}(\theta_b^a - \beta)\langle W_b^\perp, W_b^\perp\rangle - g(1 + Kn^{\beta_0}|\beta|)(\lambda + B\mu)\} > 2g(\lambda + B\mu)) = 1 - O(\exp(-cn^\epsilon))$$

Note that $\sigma_b^{-2}\langle W_b^\perp, W_b^\perp\rangle$ follows a $\chi^2_{n-|ne_a|}$-distribution. As $|ne_a| = o(n)$ and $\sigma_b^2 \geq v^2$ (by Assumption 2), it follows that there exists some $k > 0$ so that for $n > n_0$ with some $n_0(k) \in \mathbb{N}$, and any $c > 0$,

$$P(2n^{-1}\langle W_b^\perp, W_b^\perp \rangle > k) = 1 - O(\exp(-cn^\epsilon))$$

Hence, it suffices to show that for every $k, l > 0$, there exists some $n_0(k,l) \in \mathbb{N}$ so that for all $n \geq n_0$,

$$\inf_{\beta \leq 0} \{(\theta_b^a - \beta)k - l(1 + Kn^{\beta_0}|\beta|)(\lambda + B\mu)\} > 0$$

By assumption 5, $|\pi_{ab}|$ is of the order a least $n^{-\frac{1-\xi}{2}+\beta_0}$. Using

$$\pi_{ab} = \theta_b^a / (\text{Var}(X_a | X_{\Gamma(n) \setminus \{a\}}) \text{Var}(X_b | X_{\Gamma(n) \setminus \{b\}}))^{\frac{1}{2}}$$

and assumption 2, this implies that there exists some $q > 0$ so that $\theta_b^a \geq qn^{-\frac{1-\xi}{2}+\beta_0}$. As $\lambda \sim d_1 n^{-\frac{1-\varepsilon}{2}}$ and $\mu \sim d_2 n^{-\frac{1-\varepsilon}{2}-\beta_0}$ and $\xi > \varepsilon$ by assumption of theorem 1, it follows that for every $k, l > 0$ and large enough values of $n$,

$$\theta_b^a k - lKn^{\beta_0}|\beta|(\lambda + B\mu) > 0$$

It remains to show that for any $k, l > 0$, there exists some $n_0(k, l)$ such that for all $n \geq n_0$,

$$\inf_{\beta \leq 0} \{-\beta k - l(\lambda + B\mu)\} \geq 0$$

This follows as $\lambda + B\mu \to 0$ for $n \to \infty$, which completes the proof. □

**Lemma 0.1** (Accessory 1). *Assume the conditions of theorem 1 hold true. Let $R_a^{\lambda,\mu}(\beta)$ and $W_b^{\|}$ be defined as in the proof of the previous lemma. For any $g > 0$, there exists $c > 0$ so that it holds for all $a, b \in \Gamma(n)$, for $n \to \infty$,*

$$P\left(\sup_{\beta \in \mathbb{R}} \frac{|2n^{-1}\langle R_a^{\lambda,\mu}(\beta), W_b^{\|}\rangle|}{1 + Kn^{\beta_0}|\beta|} < g(\lambda + B\mu)\right) = 1 - O(\exp(-cn^\varepsilon))$$

*Proof.* By Cauchy-Schwarz inequality,

$$\frac{|2n^{-1}\langle R_a^{\lambda,\mu}(\beta), W_b^{\|}\rangle|}{1 + Kn^{\beta_0}|\beta|} \leq 2n^{-\frac{1}{2}} \| W_b^{\|} \|_2 \frac{n^{-\frac{1}{2}}\|R_a^{\lambda,\mu}(\beta)\|_2}{1 + Kn^{\beta_0}|\beta|}$$

The sum of squares of the residuals is increasing with increasing value of $\lambda, \mu$. Thus, $\| R_a^{\lambda,\mu}(\beta) \|_2 \leq \| R_a^{\infty,\infty}(\beta) \|_2$. By definition of $R_a^{\lambda,\mu}$,

$$\| R_a^{\infty,\infty}(\beta) \|_2^2 = \| X_a - \beta X_b - \beta X_{b_1} - \beta X_{b_2} - \cdots - \beta X_{b_w} \|_2^2$$

where $w$ is the number of nodes which are in the same equivalence class as $b$. And hence,

$$\| R_a^{\infty,\infty}(\beta) \|_2^2 \leq (1 + (w+1)|\beta|)^2 \max\{\| X_a \|_2^2, \| X_b \|_2^2, \| X_{b_1} \|_2^2, \cdots, \| X_{b_w} \|_2^2\} \leq (1 + Kn^{\beta_0}|\beta|)^2 \max\{\| X_a \|_2^2, \| X_b \|_2^2, \| X_{b_1} \|_2^2, \cdots, \| X_{b_w} \|_2^2\}$$

Hence, for any $q > 0$,

$$P\left(\sup_{\beta \in \mathbb{R}} \frac{n^{-\frac{1}{2}}\|R_a^{\lambda,\mu}(\beta)\|_2}{1 + Kn^{\beta_0}|\beta|} > q\right) \leq$$
$$P\left(n^{-\frac{1}{2}} \max\{\| X_a \|_2, \| X_b \|_2, \| X_{b_1} \|_2, \cdots, \| X_{b_w} \|_2\} > q\right)$$

Note that $\|X_a\|_2^2$, $\|X_b\|_2^2$ and all of $\|X_{b_k}\|_2^2$'s ($k = 1, 2, \cdots, w$) have $\chi_n^2$ distribution. Thus, by the following lemma (Lemma 0.2), there exists $q > 1$ and $c > 0$ such that

$$P\left(\sup_{\beta \in \mathbb{R}} \frac{n^{-\frac{1}{2}}\|R_a^{\lambda,\mu}(\beta)\|_2}{1 + Kn^{\beta_0}|\beta|} > q\right) = O(\exp(-cn^\varepsilon)) \text{ for } n \to \infty$$

It remains to be shown that for every $g > 0$ there exists some $c > 0$ so that

$$P\left(n^{-\frac{1}{2}}\|W_b^{\|}\|_2 > g(\lambda + B\mu)\right) = O(\exp(-cn^\varepsilon)) \text{ for } n \to \infty$$

The expression $\sigma_b^{-2}\langle W_b^{\|}, W_b^{\|}\rangle$ is $\chi_{|\text{ne}_a|-1}^2$ distributed. As $\sigma_b \leq 1$ and $|\text{ne}_a| = O(n^\kappa)$, it follows that $n^{-\frac{1}{2}}\|W_b^{\|}\|_2$ is stochastically smaller than $tn^{-\frac{1-\kappa}{2}}\left(\frac{Z}{n^\kappa}\right)^{\frac{1}{2}}$, for some $t > 0$ and for some $Z \sim \chi_{n^\kappa}^2$. Thus, for every $g > 0$,

$$P\left(n^{-\frac{1}{2}}\|W_b^{\|}\|_2 > g(\lambda + B\mu)\right) \leq P\left(\frac{Z}{n^\kappa} > (\frac{g}{t})^2 n^{1-\kappa}(\lambda + B\mu)^2\right)$$

As $\lambda \sim n^{-\frac{1-\varepsilon}{2}}$ and $\mu \sim n^{-\frac{1-\varepsilon}{2} - \beta_0}$ and $B \sim n^{\beta_0}$, it follows that $n^{1-\kappa}(\lambda + B\mu)^2 \geq hn^{\varepsilon-\kappa}$ for some $h > 0$ and sufficiently large $n$. By the properties of $\chi^2$ distribution and $\varepsilon > \kappa$, by assumption in Theorem 4.6, the claim follows. This completes the proof. □

**Lemma 0.2** (Accessory 2). *If $Y \sim \chi_n^2$, then $P(n^{-\frac{1}{2}}\sqrt{Y} > q) = O(\exp(-cn^\varepsilon))$ for some $q > 1$ and all $c > 0$*

*Proof.*

$$P(n^{-\frac{1}{2}}\sqrt{Y} > q) = P(n^{-1}Y > q^2)$$
$$= P(\frac{1}{n}\sum_{j=1}^{n} Y_j^2 > q^2) \leq \frac{\left(E\left(\exp(\frac{t}{n}Y_j^2)\right)\right)^n}{\exp(tq^2)}$$

By using Markov inequality with the increasing function $\psi_t(x) = \exp(tx)$. The moment generating function of a $\chi_1^2$ variable is known to be $\psi(s) = (1-2s)^{-\frac{1}{2}}$ for $0 \leq s < \frac{1}{2}$ and hence, the upper bound

$$P(n^{-\frac{1}{2}}\sqrt{Y} > q) \leq \exp(-tq^2)(1 - \frac{2t}{n})^{-\frac{n}{2}} = \exp\left(-tq^2 + \frac{n}{2}\log(\frac{1}{1-\frac{2t}{n}})\right)$$

Since the probability on the left-hand side does not depend on $t$, we can take the infimum over $t$ on the right (as long as the resulting $t$ satisfies the constraint $0 \leq \frac{t}{n} < \frac{1}{2}$ that is used above. We find that this infimum is achieved at $t^* = \frac{n(q^2+1)}{2}$. This $t^*$ being too large (beyond the constraint region), along with the observation that $f'(t) < 0$ for smaller values of $t$ $\left(f(t) := -tq^2 + \frac{n}{2}\log(\frac{1}{1-\frac{2t}{n}})\right)$ tells that a convenient choice for $t$ could be $\frac{n}{4}$. With this choice, we get

$$P(n^{-\frac{1}{2}}\sqrt{Y} > q) \leq \exp\left(-\frac{q^2 n}{4} + \frac{n}{2}\log 2\right)$$

And if $q^2 = \eta^2 + 2\log 2$ for $\eta > 0$ (so that $q > 1$) then this bound becomes $\leq \exp\left(-\frac{\eta^2 n}{4}\right)$. Since $\eta^2 \geq 4cn^{\varepsilon-1}$ for large $n$ and for all constant $c$, $\exp\left(-\frac{\eta^2 n}{4}\right) \leq \exp\left(-cn^{\varepsilon}\right)$. Therefore

$$P(n^{-\frac{1}{2}}\sqrt{Y} > q) = O(\exp(-cn^{\varepsilon})) \text{ for some } q > 1 \text{ and all } c > 0$$

□

**Proof of lemma 6.11.**

$$E(\epsilon_{a,i} X^a_{j,i}) = 0 \quad \forall i = 1, 2, \cdots, n$$

Also, we have

$$\frac{1}{n}\sum_{i=1}^n E\left[\left|\epsilon_{a,i} X^a_{j,i}\right|^k\right] = \frac{1}{n}\sum_{i=1}^n \left[E\left|\epsilon^k_{a,i}\right| E\left|X^a_{j,i}\right|^k\right] = E\left|\epsilon^k_{a,1}\right| E\left(\left|X^a_{j,1}\right|^k\right)$$

Using the fact that if $Z \sim N(0, \sigma^2)$, then $E(|Z|^k) = \sigma^k \cdot \frac{2^{k/2}\Gamma(\frac{k+1}{2})}{\sqrt{\pi}}$, we get the above

$$= \left[\sigma_2^k \cdot \frac{2^{k/2}\Gamma(\frac{k+1}{2})}{\sqrt{\pi}}\right] \cdot \left[(\sigma_{jj})^{k/2} \cdot \frac{2^{k/2}\Gamma(\frac{k+1}{2})}{\sqrt{\pi}}\right]$$

$$= \frac{\left(2\sigma_2\sqrt{\sigma_{jj}}\right)^k}{\pi}\left[\Gamma\left(\frac{k+1}{2}\right)\right]^2$$

$$\leq \frac{\left(2\sigma_2\sqrt{\sigma_{jj}}\right)^k}{\pi} \frac{2^{2-k}}{k+1}\Gamma(k+1) = \frac{4\sigma_2^2 \sigma_{jj}}{\pi(k+1)}\left(\sigma_2\sqrt{\sigma_{jj}}\right)^{k-2} k!$$

Under the assumption that all the population variances are 1, we get the above

$$\leq \frac{k!}{2}$$

We assumed $\sigma_2^2 = \sigma_{aa} - \Sigma_{ab}\Sigma_{bb}^{-1}\Sigma'_{ab}$ and used the following facts

- $\beta\left(\frac{k+1}{2}, \frac{k+1}{2}\right) = \frac{\Gamma\left(\frac{k+1}{2}\right)^2}{\Gamma(k+1)}$
- $\beta(x, x) = 2^{1-2x}\beta\left(x, \frac{1}{2}\right)$
- If $(a-1)(b-1) \leq 0$ then $\beta(a, b) \leq \frac{1}{ab}$

Therefore, using Bernstein's inequality, we get for $t > 0$

$$P\left(\frac{1}{n}\sum_{i=1}^n \epsilon_{a,i} X^a_{j,i} \geq t + \sqrt{2t}\right) \leq \exp(-nt)$$

Hence

$$P(\Lambda_a^c) = P\left[\max_{\substack{1 \le j \le p \\ j \ne a}} \frac{2}{n}\left|\epsilon_a' X_j^a\right| > \lambda_0 + B\mu_0\right]$$

$$= \sum_{\substack{j=1 \\ j \ne a}}^{p} P\left[\left|\frac{2}{n}\sum_{i=1}^{n} \epsilon_{a,i}' X_{j,i}^a\right| > \lambda_0 + B\mu_0\right]$$

$$= 2\sum_{\substack{j=1 \\ j \ne a}}^{p} P\left[\frac{1}{n}\sum_{i=1}^{n} \epsilon_{a,i}' X_{j,i}^a > \frac{\lambda_0 + B\mu_0}{2}\right]$$

If we choose $t = \frac{\lambda_0 + B\mu_0}{2} + 1 - \sqrt{\lambda_0 + B\mu_0 + 1}$ then we get $t + \sqrt{2t} = \frac{\lambda_0 + B\mu_0}{2}$. Therefore, using the above result, we get

$$P(\Lambda_a^c) \le 2(p-1)\exp\left[-n\left(\frac{\lambda_0 + B\mu_0}{2} + 1 - \sqrt{\lambda_0 + B\mu_0 + 1}\right)\right]$$

$$\le 2\exp\left[\log p - n\left(\frac{\lambda_0 + B\mu_0}{2} + 1 - \sqrt{\lambda_0 + B\mu_0 + 1}\right)\right]$$

$$\le 2\exp\left[\log p - n\left(\frac{\lambda_0}{2} + 1 - \sqrt{\lambda_0 + B\mu_0 + 1}\right)\right]$$

Alternatively, if we take $\lambda_0 = \frac{2(t + \log p)}{n}$ and $\mu_0 = \frac{2}{B}\sqrt{\frac{2}{n}(t + \log p)}$, we get

$$P(\Lambda_a^c) \le 2\exp(-t)$$

□




\begin{thebibliography}{10}
 
 	\bibitem{Banerjee2008}
	Banerjee, O., Ghaoui, L. E., d'Aspremont, A. (2008). ``Model Selection Through Sparse Maximum Likelihood Estimation for Multivariate Gaussian or Binary Data", \emph{Journal of Machine Learning Research}, Vol. 9, pp. 485-516.
	
	\bibitem{Beck2009}
	 Beck, A. and Teboulle, M. (2009). ``A Fast Iterative Shrinkage-thresholding Algorithm for Linear Inverse Problems". \emph{SIAM Journal on Imaging Sciences}, Vol. 2, pp. 183-202.
	 
	 \bibitem{Bickel2009}
	 Bickel, P., Ritov, Y. and Tsybakov, A. (2009). ``Simultaneous Analysis of Lasso and Dantzig Selector", \emph{Annals of Statistics}, Vol. 37, No. 4, pp. 1705-1732.
	 
	 \bibitem{Buhl1993}
	Buhl, S$\varphi$ren L. (1993), ``On the Existence of Maximum Likelihood Estimators for Graphical Gaussian Models", \emph{Scandinavian Journal of Statistics}, Vol. 20, No. 3 , pp. 263-270
	
	 \bibitem{Buhlmann2011}
	 B\"{u}hlmann, P. and van de Geer, S. (2011). ``Statistics for High-Dimensional Data". \emph{Springer Series in Statistics}.

	\bibitem{Chen2010}
	 Chen, X., Lin, Q., Kim, S., Carbonell, J. G., Xing, E. P. (2010). ``An Efficient Proximal Gradient Method for General Structured Sparse Learning", \emph{Journal of Machine Learning Research}, Vol. 11.
	 
	 \bibitem{Chen2012}
	 Chen, X., Lin, Q., Kim, S., Carbonell, J.G., and Xing, E.P. (2012). ``Smoothing Proximal Gradient Method for General Structured Sparse Regression". \emph{Annals of Applied Statistics}, Vol. 6, No. 2, pp. 719-752.
	 
	\bibitem{Dempster1972}
	Dempster, A. P. (1972). ``Covariance Selection", \emph{Biometrics}, Vol. 28, No. 1, Special Multivariate Issue, pp. 157-175.

	\bibitem{Edwards2000}
	Edwards, David (2000). ``Introduction to Graphical Modelling", Second Edition, \emph{Springer}.
	
	\bibitem{Friedman2007}
	 Friedman, J., Hastie, T. and Tibshirani, R. (2007)``Sparse inverse covariance estimation with the graphical lasso". \emph{Biostatistics}, Dec. 12, 2007,  pp. 1-10.

	\bibitem{Greenshtein2004}
	 Greenshtein, E. and Ritov, Y. (2004). ``Persistence in High-Dimensional Linear Predictor Selection and the Virtue of Overparametrization", \emph{Bernoulli}, Vol. 10, No. 6, pp. 971-988.
	
	\bibitem{Hestenes1969}
	 Hestenes, M.R. (1969). ``Multiplier and gradient methods". \emph{Journal Optimization Theory \& Applications}, Vol. 4, pp. 303-320.
	
	 \bibitem{Hoefling2010}
	 Hoefling, H. (2010). ``A Path Algorithm for the Fused Lasso Signal Approximator". \emph{Journal of
Computational and Graphical Statistics}, Vol. 19, No. 4, pp. 984-1006.

	\bibitem{Honorio2009}
	Honorio, Jean, Ortiz, Luis, Samaras, Dimitris, Paragios, Nikos, Goldstein, Rita (2009), ``Sparse and Locally Constant Gaussian Graphical Models", \emph{Advances in Neural Information Processing Systems 22}.

	\bibitem{Kovac2012}
	 Kovac, A. and Smith, A. D. A. C. (2012), ``Nonparametric Regression on a Graph". \emph{Journal of Computational and Graphical Statistics}. Vol. 20, No. 2, pp. 432-447.
	 
	\bibitem{Lauritzen1996}
	Lauritzen, Stephen L. (1996), ``Graphical Models". \emph{Oxford Statistical Science Series: Clarendon Press, Oxford}.
	
	 \bibitem{Liu2010}
	 Liu, J., Yuan, L., and Ye, J. (2010). ``An Efficient Algorithm for a Class of Fused Lasso Problems".
In The \emph{ACM SIG Knowledge Discovery and Data Mining}. ACM, Washington, DC.
	 
	 \bibitem{Meinshausen2006}
	Meinshausen, N. and B\"{u}hlmann, P. (2006). ``High-Dimensional Graphs and Variable Selection with the LASSO", \emph{The Annals of Statistics}, Vol. 34, No. 3, pp. 1436-1462.

	\bibitem{Nesterov2005}
	 Nesterov, Y. (2005). ``Smooth Minimization of Non-smooth Functions". \emph{Mathematical Programming}, Vol. 103, pp. 127-152.
	 
	 \bibitem{Rockafeller1973}
	 Rockafellar, R.T. (1973), ``A Dual Approach to Solving Nonlinear Programming Problems by
Unconstrained Optimization". \emph{Mathematical Programming}, Vol. 5, pp. 354-373.
	 
	\bibitem{Speed1986}
	Speed, T. P. and Kiiveri, H. T., ``Gaussian Markov Distributions over Finite Graphs", \emph{The Annals of Statistics}, Vol. 14, No. 1, pp. 138-150.
	
	\bibitem{Tibshirani1997}
	Tibshirani, Robert (1997). ``Regression Shrinkage and Selection via the Lasso", \emph{Journal of Royal Statistical Society}, Series B (Methodological), Vol. 58, pp. 267-288.
                
        \bibitem{Tibshirani2005}
	Tibshirani, R., Saunders, M., Rosset, S., Zhu, Ji and Knight, K. (2005). ``Sparsity and Smoothness via the Fused Lasso", \emph{Journal of Royal Statistical Society}, Series B (Methodological), Vol. 67, Part 1, pp. 91-108.
	
	\bibitem{Tibshirani2011}
	 Tibshirani, R.J. and Taylor, J. (2011). ``The Solution Path of the Generalized Lasso". \emph{The
Annals of Statistics}, Vol. 39, No. 3, pp. 1335-1371.
	 
	\bibitem{Uhler2012}
	Uhler, Caroline (2012), ``Geometry of Maximum Likelihood Estimation in Gaussian Graphical Models", \emph{The Annals of Statistics}, Vol. 40, No. 1, 238-261.
	
	\bibitem{Vandenberghe1998}
	Vandenberghe, L., Boyd, S. and Wu, S.-P. (1998). ``Determinant Maximization with Linear Matrix Inequality Constraints", \emph{SIAM Journal on Matrix Analysis and Applications}, Vol. 19, No. 2, pp. 499-533.
	
	 \bibitem{Ye2011}
	 Ye, G.B. and Xie, X. (2011). ``Split Bregman Method for Large Scale Fused Lasso". \emph{Computational Statistics \& Data Analysis}, Vol. 55, No. 4, pp. 1552-1569.

	\bibitem{Yuan2007}
	Yuan, M. and Lin, Y. (2007). ``Model Selection and Estimation in the Gaussian Graphical Model", \emph{Biometrika}, Vol. 94, No. 1, pp. 19-35.
	
\end{thebibliography}
\end{document}